\documentclass{LMCS}

\def\doi{8 (2:16) 2012}
\lmcsheading%
{\doi}
{1--32}
{}
{}
{Oct.~12, 2011}
{Jun.~26, 2012}
{}

\usepackage{enumerate}
\usepackage{hyperref}
\usepackage{color}
\usepackage{epsfig}
\usepackage{pstricks}
\usepackage{latexsym}
\usepackage{amssymb}
\usepackage{amsmath}
\usepackage{algorithmic}
\usepackage[ruled]{algorithm}
\usepackage{calc}
\usepackage{nicefrac}
\usepackage{wasysym}

\newcommand\hiddencomment[1]{}

\newcommand{\Nat}{{\mathbb N}}

\newcommand{\Bool}{{\mathbb B}}

\newcommand\false{{\tt false}}
\newcommand\true{{\tt true}}

\renewcommand\implies{\,\Rightarrow\,}

\newcommand{\vect}[1]{\boldsymbol{#1}}

\newcommand\algrule[2]{\frac{
    \begin{array}{c}
      #1
    \end{array}
  }{#2}}

\newcommand\Trans{{\it Trans}}

\newcommand\Init{{\it Init}}

\newcommand\Target{{\it Target}}
\newcommand\Region{{\it Region}}

\newcommand\BReach{{\B}}

\newcommand\B{{\mathcal B}}
\newcommand\Reg{{\mathcal R}}
\newcommand\Kernel{{\mathcal K}}

\newcommand\MaxReach{{\it MaxReach}}
\newcommand\MinStable{{\it MinStable}}
\newcommand\MinReach{{\it MinReach}}
\newcommand\MaxAvoid{{\it MaxAvoid}}

\newcommand\ff{{\it ff}}

\newrgbcolor{darkgreen}{0 0.6 0}
\newrgbcolor{saddlebrown}{0.55 0.27 0.07}

\newcounter{rulecounter}
\newenvironment{RULE}
{%
\refstepcounter{rulecounter}%
\addtocounter{equation}{-1} 
\begin{equation}}
{\end{equation}}

\newcounter{rulecounter2}
\newenvironment{RULE2}
{%
\refstepcounter{rulecounter2}%
\addtocounter{equation}{-1} 
\begin{equation}}
{\end{equation}}

\newcommand{\sol}{{\mathcal S}_{A,B}}
\newcommand{\MDP}{{\mathcal M}}

\newcommand{\R}{\rotatebox[origin=c]{180}{{\sf R}}}

\newcommand{\Q}{{\mathcal Q}}
\newcommand{\Int}{{\mathcal I}}

\allowdisplaybreaks[4]

\begin{document}

\title[Generalized Craig Interpolation for SSAT]{Generalized Craig Interpolation
for Stochastic Boolean Satisfiability Problems with Applications to
Probabilistic State Reachability and Region Stability}

\author[T.~Teige]{Tino Teige}   
\address{Carl von Ossietzky University of Oldenburg\\
Department of Computing Science\\
Research Group Hybrid Systems\\
D-26111 Oldenburg, Germany} 
\email{\{tino.teige,fraenzle\}@informatik.uni-oldenburg.de}  
\thanks{This work has been supported by the German Research Council (DFG) as
part of the Trans\-regional Collaborative Research Center ``Automatic
Verification and Analysis of Complex Systems'' (SFB/TR 14 AVACS,
\url{www.avacs.org})
as well as by the European Union Seventh Framework Programme FP7/2007-2013
under the MoVeS Project (grant agreement No.~257005,
\url{http://www.movesproject.eu}).}

\author[M.~Fr{\"a}nzle]{Martin Fr{\"a}nzle} 
\address{\vskip-6 pt} 





\begin{abstract}
\noindent The stochastic Boolean satisfiability (SSAT) problem has been
introduced by Papadimitriou in 1985 when adding a probabilistic model of
uncertainty to propositional satisfiability through randomized quantification.
SSAT has many applications, among them probabilistic bounded model checking
(PBMC) of symbolically represented Markov decision processes. This article
identifies a notion of \emph{Craig interpolant} for the SSAT framework and
develops an algorithm for computing such interpolants based on a resolution
calculus for SSAT.

As a potential application area of this novel concept of Craig interpolation,
we address the symbolic analysis of probabilistic systems. We first investigate
the use of interpolation in probabilistic state reachability analysis, turning
the falsification procedure employing PBMC into a verification technique for
probabilistic safety properties.
We furthermore propose an interpolation-based approach to probabilistic region
stability, being able to verify that the probability of stabilizing within some
region is sufficiently large.
\end{abstract}

\maketitle

\section*{Introduction}
\label{Sec:Intro}

Papadimitriou \cite{Papadimitriou85} has proposed the idea of modeling
uncertainty within propositional satisfiability (SAT) by adding
\emph{randomized} quantification to the problem description. The resultant
\emph{stochastic Boolean satisfiability} (SSAT) problems consist of a
quantifier prefix followed by a propositional formula. The quantifier
prefix is an alternating sequence of existentially quantified
variables and variables bound by randomized quantifiers. The meaning
of a randomized variable $x$ is that $x$ takes value $\true$ with a
certain probability $p$ and value $\false$ with the complementary
probability $1-p$. Due to the presence of such probabilistic
assignments, the semantics of an SSAT formula $\Phi$ no longer is
qualitative in the sense that $\Phi$ is satisfiable or unsatisfiable,
but rather \emph{quantitative} in the sense that we are interested in
the \emph{maximum probability of satisfaction} of $\Phi$.
Intuitively, a solution of $\Phi$ is a strategy for assigning the
existential variables, i.e.\ a tree of assignments to the
existential variables depending on the probabilistically determined
values of preceding randomized variables, such that the assignments
maximize the probability of satisfying the propositional formula.

In recent years, the SSAT framework has attracted interest within the
Artificial Intelligence community, as many problems from that area involving
uncertainty have concise descriptions as SSAT problems, in particular
probabilistic planning problems~%
\cite{littman01stochastic,majercik98maxplan,majercik03contingent}.
Inspired by that work, other communities have started to exploit SSAT
and closely related formalisms within their domains. The Constraint
Programming community is working on \emph{stochastic constraint
satisfaction} problems~%
\cite{walsh02stochastic,BalafoutisS06} to address, among others,
multi-objective decision making under uncertainty~\cite{BordeauxS07}.
Recently, a technique for the symbolic analysis of probabilistic
hybrid systems based on stochastic satisfiability has been suggested
by the authors~%
\cite{FraHerTei08:SSMT,TeigeFraenzle09:ADHS,FraTeiEgg:JLAP,TeiEggFra:NAHS}.
To this end, SSAT has been extended by embedded theory reasoning over arithmetic
theories, as known from \emph{satisfiability modulo theories}
(SMT)~\cite{BSST09HBSAT}, which yields the notion of \emph{stochastic
satisfiability modulo theories} (SSMT). By the expressive power of
SSMT, bounded probabilistic reachability problems of uncertain hybrid
systems can be phrased symbolically as SSMT formulae yielding the same
probability of satisfaction~%
\cite{FraHerTei08:SSMT,TeigeFraenzle09:ADHS,FraTeiEgg:JLAP,TeiEggFra:NAHS}.
As this bounded model checking approach yields valid lower bounds $lb$ of the
probability of reaching undesirable system states along unbounded runs, it is
able to \emph{falsify} probabilistic safety requirements of shape ``a system
error occurs with probability at most $0.1\permil$'', namely if a lower bound
$lb > 0.1\permil$ is computed.

Though the general SSAT problem and even its restriction to 2CNF,
i.e.\ to formulae in conjunctive normal form containing clauses with
two literals only, are {\rm PSPACE}-complete~\cite{TeiFrae:LPAR17},
the plethora of real-world applications calls for practically
efficient algorithms.  The first SSAT algorithm, suggested by
Littman~\cite{littman99initial}, extends the
Davis-Putnam-Logemann-Loveland (DPLL) procedure~%
\cite{DavisPutnam,DavisEA:DPLL62} for SAT with appropriate quantifier
handling and algorithmic optimizations like \emph{thresholding}.
Majercik further improved the DPLL-based SSAT algorithm by
\emph{non-chronological
  backtracking}~\cite{Majercik04Nonchronological}.  The SSMT algorithm
from~\cite{FraHerTei08:SSMT,TeiFra08:nonlinearSSMT,TeigeFraenzle09:ADHS,
  TeiEggFra:NAHS} being implemented in the SSMT tool SiSAT builds on
the DPLL-based SSAT procedures plus conflict-driven clause learning,
but also integrates an underlying theory solver addressing non-linear
arithmetics, and was successfully applied to realistic case studies
featuring hybrid discrete-continuous state
spaces~\cite{TeigeFraenzle09:ADHS,FraTeiEgg:JLAP,TeiEggFra:NAHS}.
Unlike these explicit tree-traversal approaches and motivated by work
on \emph{resolution} for propositional and first-order
formulae~\cite{Robinson:Resolution65} and for quantified Boolean
formulae (QBF)~\cite{BuningKF95}, the authors have recently developed
an alternative SSAT procedure based on
resolution~\cite{TeiFrae:LPAR17}.

In this article, we investigate the concept of Craig interpolation for
SSAT. Given two formulae $A$ and $B$ for which $A \implies B$ is true,
a \emph{Craig interpolant}~\cite{Craig57} $\Int$ is a formula over
variables common to $A$ and $B$ that ``lies in between'' $A$ and $B$
in the sense that $A \implies \Int$ and $\Int \implies B$. In the
automatic hardware and software verification communities, Craig
interpolation has found widespread use in model checking algorithms,
both as a means of extracting reasons for non-concretizability of a
counterexample obtained on an abstraction as well as for obtaining a
symbolic description of reachable state sets. In McMillan's
approach~\cite{McMillan03,McMillan:TACAS05}, interpolants are used to
symbolically describe an overapproximation of the step-bounded
reachable state set. If the sequence of interpolants thus obtained
stabilizes eventually, i.e.\ no additional state is found to be
reachable, then the corresponding state-set predicate $R$ has all
reachable system states as its models. The safety property that states
satisfying $B$, where $B$ is a predicate, are never reachable is then
verified by checking $R \wedge B$ for unsatisfiability.

Given McMillan's verification approach to reachability analysis of
non-probabilistic systems based on Craig interpolation for SAT, it is natural to
ask whether a corresponding probabilistic counterpart can be developed, i.e.\ a
\emph{verification approach to probabilistic reachability analysis of
probabilistic systems based on Craig interpolation for stochastic SAT}. Such an
approach would complement the aforementioned falsification procedure for
probabilistic systems based on SSAT/SSMT. In this article, we suggest a solution
to the issue above.

In addition to probabilistic state reachability, we address the
problem of \emph{probabilistic region stability}. The latter problem
is motivated by the notion of region stability for non-probabilistic
hybrid systems~\cite{PodelskiW:FORMATS07,PodelskiW:HSCC07}, where a
system is called stable with respect to some region $R$ iff all system
runs eventually reach $R$ and finally stay in $R$ forever. In this
article, we suggest an adaptation of region stability to the
probabilistic case along with a symbolic, interpolation-based
procedure for the verification of probabilistic stability properties
like ``the probability that the system stabilizes within region $R$ is
at least $99.9\%$''.

\paragraph{Structure of the article.}
After a formal introduction to SSAT in Section~\ref{Sec:Prelim},
Section~\ref{Sec:GCIdef} is devoted to a generalization of the notion of Craig
interpolants suitable for SSAT. Thereafter, Section~\ref{Sec:GCIcompute}
elaborates on an algorithm for computing such generalized Craig interpolants,
which relies on a resolution calculus for SSAT. The application of generalized
Craig interpolation to the symbolic analysis of probabilistic systems, namely to
probabilistic state reachability as well as to probabilistic region stability,
is then addressed in Section~\ref{Sec:Analysis}, where applicability of these
novel techniques is illustrated on small examples. Section~\ref{Sec:Conclusion}
finally concludes the article.

\section{Stochastic Boolean satisfiability}
\label{Sec:Prelim}

\begin{figure}[t]
\centering\resizebox{0.5\textwidth}{!}
{\input{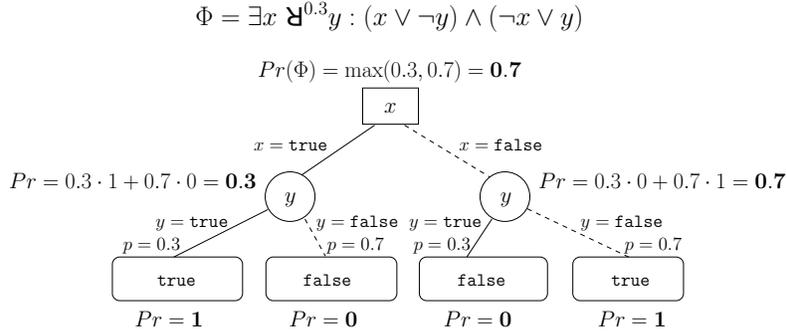}}\vspace*{-1.5mm}
\caption{Semantics of an SSAT formula depicted as a tree.}
\label{Fig:SSAT_semantics}
\end{figure}
A \emph{stochastic Boolean satisfiability} (SSAT) formula is of the form
$\Phi = \Q : \varphi$ with a prefix $\Q = Q_1 x_1 \ldots Q_n x_n$ of quantified
propositional variables $x_i$, where $Q_i$ is either an existential quantifier
$\exists$ or a randomized quantifier $\R^{p_i}$ with a rational constant $0 <
p_i < 1$, and a propositional formula $\varphi$ such that ${\it Var}(\varphi)
\subseteq \{x_1, \ldots, x_n\}$, where ${\it Var}(\varphi)$ denotes the set of
all (necessarily free) variables occurring in $\varphi$.
Note that SSAT formula $\Phi$ thus has no free variables.
Without loss of generality, we assume that $\varphi$ is in \emph{conjunctive
normal form} (CNF), i.e.\ a conjunction of disjunctions of propositional
literals. A \emph{literal} $\ell$ is a propositional variable, i.e.\
$\ell=x_i$, or its negation, i.e.\ $\ell = \neg x_i$. A \emph{clause}
is a disjunction of literals. Throughout the article and without loss of
generality, we require that a clause does not contain the same literal more than
once as $\ell \vee \ell \equiv \ell$. Consequently, we may also identify a
clause with its set of literals.
The semantics of $\Phi$, as illustrated in Figure~\ref{Fig:SSAT_semantics}, is
defined by the \emph{maximum probability of satisfaction} $Pr(\Phi)$ as follows.
\[
\begin{array}{l@{~=~}l}
Pr(\varepsilon : \varphi) &
\left\{
\begin{array}{l}
0 \text{~~if~} \varphi \text{ is logically equivalent to \false}\\
1 \text{~~if~} \varphi \text{ is logically equivalent to \true}
\end{array}
\right.\\
Pr(\exists x\ \Q : \varphi) &
\max(Pr(\Q : \varphi[\true/x]), Pr(\Q : \varphi[\false/x]))\\
Pr(\R^{p} x\ \Q : \varphi) &
p \cdot Pr(\Q : \varphi[\true/x])\ +\ (1-p) \cdot Pr(\Q : \varphi[\false/x])\\
\end{array}
\]
Note that the semantics is well-defined as $\Phi$ has no free
variables such that all variables have been substituted by the constants $\true$
and $\false$ when reaching the quantifier-free base case.



\section{Generalized Craig interpolants}
\label{Sec:GCIdef}

Craig interpolation~\cite{Craig57} is a well-studied notion in formal
logics which has several applications in Computer Science, among them
model checking~\cite{McMillan03,McMillan:TACAS05}. Given two formulae $\varphi$
and $\psi$ such that $\varphi \implies \psi$ is valid, a \emph{Craig
interpolant} for $(\varphi,\psi)$ is a formula $\Int$ which refers only
to common variables of $\varphi$ and $\psi$, and $\Int$ is
``intermediate'' in the sense that $\varphi \implies \Int$ and $\Int
\implies \psi$. Such interpolants do trivially exist in all logics
permitting quantifier elimination, for instance, in propositional logic.
The observation that $\varphi \implies \psi$ holds iff $\varphi \wedge
\neg \psi$ is unsatisfiable gives rise to an equivalent definition which
we refer to in the rest of the article:\footnote{This is of technical nature
as SSAT formulae are interpreted by the maximum probability of
satisfaction. As the \emph{maximum} probability that an implication
$\varphi \implies \psi$ holds is inappropriate for our purpose, we
reason about the maximum satisfaction probability $p$ of the negated
implication, i.e.\ of $\varphi \wedge \neg \psi$, instead.  The
latter relates to the \emph{minimum} probability $1-p$ that
$\varphi \implies \psi$ holds, which is the desired notion.} given
an unsatisfiable formula $\varphi \wedge \neg\psi$, a formula $\Int$ is a Craig
interpolant for $(\varphi, \psi)$ iff both $\varphi \wedge \neg
\Int$ and $\Int \wedge \neg\psi$ are unsatisfiable and $\Int$ mentions
only common variables.

In this section, we investigate the issue of Craig interpolation for stochastic
SAT. We propose a generalization of Craig interpolants suitable for SSAT and
show the general existence of such interpolants.
In Section~\ref{Sec:GCIcompute}, we then devote our attention to an automatic
method for computing generalized Craig interpolants based on a resolution
calculus for SSAT.


When approaching a reasonable definition of interpolants for SSAT, the
semantics of the non-classical quantifier prefix poses problems: Let
$\Phi = \Q: (A \wedge B)$ be an SSAT formula. Each variable in $A
\wedge B$ is bound by $\Q$, which provides the probabilistic
interpretation of the variables that is lacking without the quantifier
prefix. This issue can be addressed by considering the quantifier
prefix $\Q$ as the global setting that serves to interpret the
quantifier-free part, and consequently interpreting the interpolant
also within the scope of $\Q$, thus reasoning about $\Q: (A \wedge
\neg \Int)$ and $\Q: (\Int \wedge B)$.  A more fundamental problem is
that a classical Craig interpolant for $\Phi$ only exists if
$Pr(\Phi)=0$, since $A \wedge B$ has to be unsatisfiable by definition
of a Craig interpolant which applies iff $Pr(\Q: (A \wedge B))=0$. The
precondition that $Pr(\Q: (A \wedge B))=0$ would be far too
restrictive for application of interpolation, as the notion of
unsatisfiability of $A \wedge B$ is naturally generalized to
satisfiability with insufficient probability, i.e.\ $Pr(\Q: (A \wedge
B))$ being ``sufficiently small'', in the stochastic setting. Such
relaxed requirements actually appear in practice, for instance, in
probabilistic verification where safety properties like ``a fatal
system error is never reachable'' are frequently replaced by
probabilistic ones like ``a fatal system error is reachable only with
(sufficiently small) probability of at most $0.1\permil$''. Motivated
by above facts, interpolants for SSAT should also exist when $A \wedge B$
is satisfiable with reasonably low probability.

The resulting notion of interpolation, which is to be made precise in
Definition~\ref{Def:GenInt}, matches the following intuition. In
classical Craig interpolation, when performed in logics permitting
quantifier elimination, the Craig interpolants of $(A,\neg B)$ form a
lattice with implication as its ordering, $A^{\exists} = \exists
a_1,\ldots a_\alpha:A$ as its bottom element and $\overline{B}^{\forall} =
\neg\exists b_1,\ldots b_\beta:B$ as its top element, where the $a_i$ and $b_i$
are the local variables of $A$ and of $B$, respectively. In the generalized
setting required for
SSAT\footnote{Though the concept seems to be more general, this article
addresses SSAT only.}, $A\Rightarrow\neg B$ and thus
$A^\exists \Rightarrow \overline{B}^\forall$ may no longer hold such that the
above lattice can collapse to the empty set. To preserve the overall
structure, it is however natural to use the lattice of propositional
formulae ``in between'' $A^\exists \wedge \overline{B}^\forall$ as bottom
element and $A^\exists \vee \overline{B}^\forall$ as top element
instead. This lattice is non-empty and coincides with the classical
one whenever $A\wedge B$ is unsatisfiable.
\begin{defi}[Generalized Craig interpolant]
\label{Def:GenInt}
Let $A,B$ be propositional formulae and
$V_A := {\it Var}(A) \setminus {\it Var}(B) = \{a_1, \ldots, a_{\alpha}\}$,
$V_B := {\it Var}(B) \setminus {\it Var}(A) = \{b_1, \ldots, b_{\beta}\}$,
$V_{A,B} := {\it Var}(A) \cap {\it Var}(B)$,
$A^{\exists} = \exists a_1, \ldots, a_{\alpha}:A$, and
$\overline{B}^\forall = \neg \exists b_1, \ldots, b_{\beta}: B$.
A propositional formula $\Int$ is called \emph{generalized Craig
interpolant for $(A,B)$} iff 
${\it Var}(\Int) \subseteq V_{A,B}$,
$\left(A^{\exists} \wedge \overline{B}^\forall\right) \implies \Int$, and
$\Int \implies \left(A^{\exists} \vee \overline{B}^\forall\right)$.
\end{defi}
Given any two propositional formulae $A$ and $B$, the four quantifier-free
propositional formulae equivalent to
$A^\exists \wedge \overline{B}^\forall$,
to $A^\exists$,
to $\overline{B}^\forall$, and
to $A^\exists \vee \overline{B}^\forall$,
are generalized Craig interpolants for $(A,B)$. These generalized interpolants
always exist since propositional logic has quantifier elimination.

\begin{figure}[t]
\centering\resizebox{0.9\textwidth}{!}
{\input{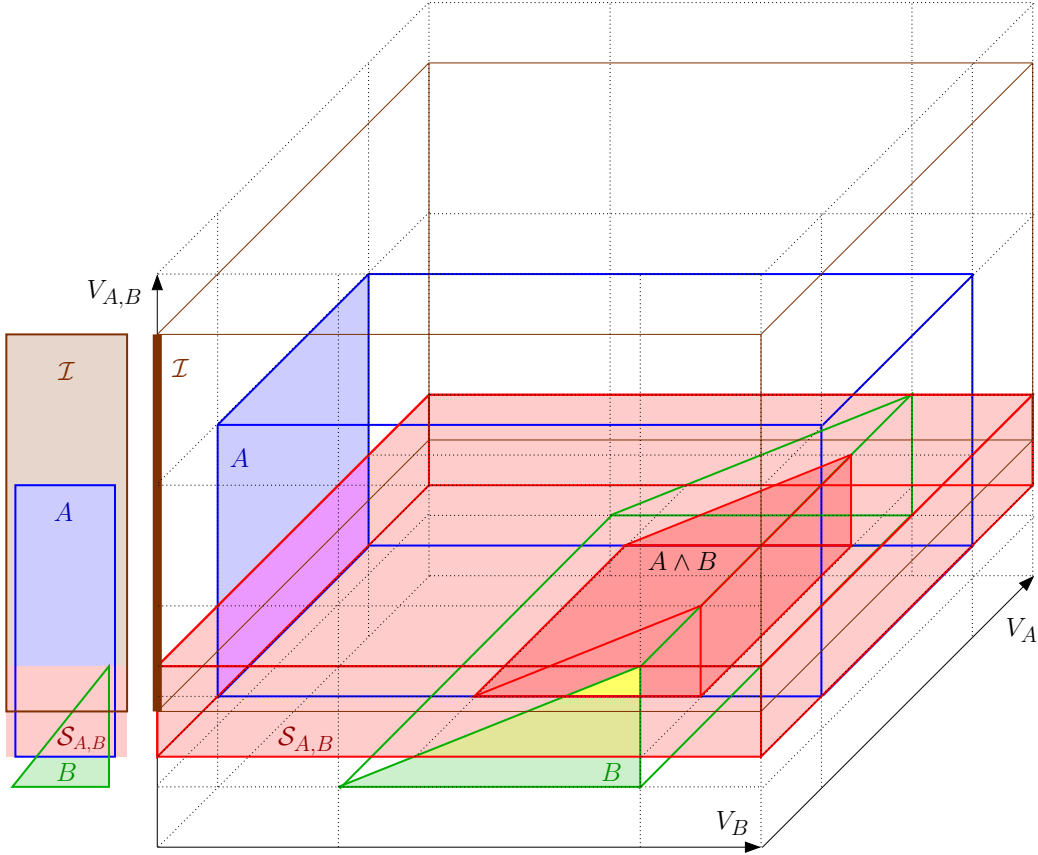}}\vspace*{-1.5mm}
\caption{Geometric interpretation of a generalized Craig interpolant
  $\Int$. $V_A$-, $V_B$-, and $V_{A,B}$-axes denote assignments of
  variables occurring only in $A$, only in $B$, and in both $A$ and
  $B$, respectively.}
\label{Fig:stoch-interpolant}
\end{figure}

While Definition~\ref{Def:GenInt} motivates the generalized notion of Craig
interpolant from a model-theoretic perspective, we state an equivalent
definition of generalized Craig interpolants in Lemma~\ref{Lemma:InterSSAT}
that substantiates the intuition of generalized interpolants and allows for an
illustration of their geometric shape. Given two formulae $A$ and $B$, the idea
of generalized Craig interpolant is depicted in
Figure~\ref{Fig:stoch-interpolant}. The set of solutions of $A$ is
defined by the rectangle on the $V_A,V_{A,B}$-plane with a cylindrical
extension in $V_B$-direction as $A$ does not contain variables in
$V_B$. Similarly, the solution set of $B$ is given by the triangle on
the $V_B,V_{A,B}$-plane and its cylinder in $V_A$-direction. The
solution set of $A \wedge B$ is then determined by the intersection of
both cylinders. Since $A \wedge B \wedge \neg (A\wedge B)$ is
unsatisfiable, the sets $A \wedge \neg (A\wedge B)$ and $B \wedge \neg
(A\wedge B)$ are disjoint. This gives us the possibility to talk about
interpolants wrt.\ these sets. However, a formula $\Int$ over only common
variables in $V_{A,B}$ may not exist when demanding
$A \wedge \neg (A\wedge B) \wedge \neg \Int$ and $\Int \wedge B \wedge
\neg (A\wedge B)$ to be unsatisfiable. This is indicated by
Figure~\ref{Fig:stoch-interpolant} and proven by the simple example
$A=(a)$, $B=(b)$. As $V_{A,B}=\emptyset$, $\Int$ is
either $\true$ or $\false$. In first case, $\true \wedge (b) \wedge
\neg(a \wedge b)$ is satisfiable, while $(a) \wedge \neg
(a\wedge b) \wedge \neg \false$ is in second case.  If we however project
the solution set of $A \wedge B$ onto the $V_{A,B}$-axis and subtract
the resulting hyperplane $\sol$ from $A$ and $B$ then such a formula
$\Int$ over $V_{A,B}$-variables exists. The next lemma formalizes such
generalized interpolants $\Int$ and shows their equivalence to the ones from
Definition~\ref{Def:GenInt}.
\begin{lem}[Generalized Craig interpolant for SSAT]
\label{Lemma:InterSSAT}
Let $\Phi = \Q: (A \wedge B)$ be some SSAT formula, $V_A$, $V_B$, $V_{A,B}$
be defined as in Definition~\ref{Def:GenInt}, and $\sol$ be a propositional
formula with ${\it Var}(\sol) \subseteq V_{A,B}$ such that $\sol \equiv \exists
a_1, \ldots, a_{\alpha}, b_1, \ldots, b_{\beta}: (A \wedge B)$.
Then, a propositional formula $\Int$ is a generalized Craig
interpolant for $(A,B)$ iff the following properties are satisfied.
\begin{enumerate}[\em(1)]
\item ${\it Var}(\Int) \subseteq V_{A,B}$
\item $Pr(\Q: (A \wedge \neg \sol \wedge \neg \Int)) = 0$
\item $Pr(\Q: (\Int \wedge B \wedge \neg \sol)) = 0$
\end{enumerate}
\end{lem}
\proof
As ${\it Var}(\Int) \subseteq V_{A,B}$ holds for generalized Craig
interpolants
$\Int$, it remains to show that
$(A^{\exists} \wedge \overline{B}^\forall) \implies \Int$ and
$\Int \implies (A^{\exists} \vee \overline{B}^\forall)$ iff
$Pr(\Q: (A \wedge \neg \sol \wedge \neg \Int)) = 0$ and $Pr(\Q: (\Int \wedge B
\wedge \neg \sol)) = 0$.
Observe that
$\models (A^{\exists} \wedge \overline{B}^{\forall}) \implies \Int$ iff
$\models \forall a_1, \ldots, a_{\alpha}: (A \wedge \overline{B}^{\forall})
\implies \Int$ iff
$\models (A \wedge \overline{B}^{\forall}) \implies \Int$ iff
$\models (A \wedge (\neg A^{\exists} \vee \overline{B}^{\forall})) \implies
\Int$ iff
$\models (A \wedge \neg \sol) \implies \Int$ iff
$A \wedge \neg \sol \wedge \neg \Int$ is unsatisfiable iff $Pr(\Q: (A \wedge
\neg \sol \wedge \neg \Int)) = 0$.
Analogously, $\models \Int \implies (A^{\exists} \vee \overline{B}^{\forall})$
iff
$\models \forall b_1, \ldots, b_{\beta}: \Int \implies (A^{\exists} \vee \neg
B)$ iff $\models \Int \implies (A^{\exists} \vee \neg B)$ iff $\models \Int
\implies ((A^{\exists} \wedge \neg \overline{B}^{\forall}) \vee \neg B)$ iff
$\models \Int \implies (\sol \vee \neg B)$ iff $\Int \wedge \neg \sol \wedge
B$
is unsatisfiable iff $Pr(\Q: (\Int \wedge B \wedge \neg \sol)) =
0$.\qed
We remark that the concept of generalized Craig interpolants is a
generalization of Craig interpolants in the sense that whenever $A
\wedge B$ is unsatisfiable, i.e.\ when $Pr(\Q: (A \wedge B)) = 0$,
then each generalized Craig interpolant $\Int$ for $(A,B)$ actually is
a Craig interpolant for $A$ and $B$ since $\sol \equiv \false$.

\section{Computation of generalized Craig interpolants}
\label{Sec:GCIcompute}

In this section, we proceed to the efficient computation of
generalized Craig interpolants. The remark following
Definition~\ref{Def:GenInt} shows that generalized interpolants can in
principle be computed by \emph{explicit} quantifier elimination methods, like
Shannon's expansion or binary decision diagrams (BDDs). We aim at a more
efficient method based on SSAT resolution~\cite{TeiFrae:LPAR17} akin to
resolution-based Craig interpolation for propositional SAT by
Pudl\'{a}k~\cite{Pudlak97}. The latter approach has been integrated into
DPLL-based SAT solvers featuring conflict analysis and successfully applied to
symbolic model checking~\cite{McMillan03,McMillan:TACAS05}.
%
To this end, we first recall the sound and complete resolution calculus for SSAT
from~\cite{TeiFrae:LPAR17} in Section~\ref{Subec:S-Resol}. Thereafter, SSAT
resolution is enhanced in order to compute generalized Craig interpolants in
Section~\ref{Subec:InterpolatingS-Resol}.

\subsection{Resolution for SSAT}
\label{Subec:S-Resol}

As basis of the SSAT interpolation procedure introduced in
Section~\ref{Subec:InterpolatingS-Resol}, we recall the sound and complete
resolution calculus for SSAT from~\cite{TeiFrae:LPAR17}, subsequently called
\emph{S-resolution}. In contrast to SSAT algorithms implementing a
DPLL-based backtracking procedure, thereby explicitly traversing the
tree given by the quantifier prefix and recursively computing the
individual satisfaction probabilities for each subtree by the scheme
illustrated in Figure~\ref{Fig:SSAT_semantics}, S-resolution follows the
idea of \emph{resolution} for propositional and first-order
formulae~\cite{Robinson:Resolution65} and for QBF
formulae~\cite{BuningKF95} by deriving new clauses $c^p$ annotated
with probabilities $0 \leq p \leq 1$. S-resolution differs from
non-stochastic resolution, as such derived clauses $c^p$ need not be
implications of the given formula, but are just entailed with some
probability. Informally speaking, the derivation of a clause
$c^p$ means that under SSAT formula $\Q: \varphi$, the clause $c$ is
violated with a maximum probability at most $p$, i.e.\ the
satisfaction probability of $\Q: (\varphi \wedge \neg c)$ is at most
$p$.  More intuitively, the minimum probability that clause $c$ is
implied by $\varphi$ is at least $1-p$.\footnote{We remark that
$Pr(\Q: \psi) = 1 - Pr(\Q': \neg\psi)$, where $\Q'$ arises from $\Q$
by replacing existential quantifiers by universal ones, where
universal quantifiers call for \emph{minimizing} the satisfaction
probability.}  Once an annotated empty clause $\emptyset^p$ is
derived, it follows that the probability of the given SSAT formula is
at most $p$, i.e.\ $Pr(\Q: (\varphi \wedge \neg \false)) = Pr(\Q:
\varphi) \leq p$.

In what follows, let $\Q: \varphi$ be an SSAT formula with $\varphi$ in CNF.
Without loss of generality, $\varphi$ contains only non-tautological
clauses\footnote{Tautological clauses $c$, i.e.\ $\models c$,
are redundant, i.e.\ $Pr(\Q: (\varphi \wedge c)) = Pr(\Q: \varphi)$.}, i.e.\
$\forall c \in \varphi:\ \not\models c$.  Let $\Q = Q_1 x_1 \ldots Q_n
x_n$ be the quantifier prefix and $\varphi$ be some propositional
formula with ${\it Var}(\varphi) \subseteq \{x_1,\ldots,x_n\}$. The
quantifier prefix $\Q(\varphi)$ is defined to be shortest prefix of
$\Q$ that contains all variables from $\varphi$, i.e.\
$\Q(\varphi)=Q_1 x_1 \ldots Q_i x_i$ where $x_i \in {\it
Var}(\varphi)$ and for each $j > i: x_j \notin {\it Var}(\varphi)$.
Let further be ${\it Var}(\varphi)\downarrow_{k} := \{x_1, \ldots, x_k\}$ for
each integer $0 \leq k \leq n$.
For a non-tautological clause $c$, i.e.\ if $\not\models c$, we define
the unique assignment $\ff_c$ that falsifies $c$ as the mapping
\[\ff_c: {\it Var}(c) \to \Bool \text{~such that~} \forall x \in {\it Var}(c):
\ff_c(x)=\left\{
\begin{array}{l@{~;~}r}
\true   & \neg x \in c,\\
\false  & x \in c.\\
\end{array}
\right.\]
Consequently, $c$ evaluates to $\false$ under assignment $\ff_c$.

Starting with clauses in $\varphi$, \emph{S-resolution} is given by the
consecutive application of rules~\ref{Rule:conflict} to \ref{Rule:resolution} to
derive new clauses $c^p$ with $0 \leq p \leq 1$.
Rule~\ref{Rule:conflict} derives a clause $c^0$ from an original clause $c$ in
$\varphi$. Referring to the definition of $Pr(\Q:\varphi)$ in
Section~\ref{Sec:Prelim}, \ref{Rule:conflict} corresponds to the quantifier-free
base case where $\varphi$ is equivalent to $\false$ under any assignment that
falsifies $c$.
\begin{RULE}\label{Rule:conflict}
\algrule%
{%
c \in \varphi
}%
{%
c^0
}%
\end{RULE}%
Similarly, \ref{Rule:solution} reflects the quantifier-free base case in which
$\varphi$ is equivalent to $\true$ under any assignment $\tau'$ that is conform
to the partial assignment $\tau$ since $\models
\varphi[\tau(x_1)/x_1]\ldots[\tau(x_{i})/x_{i}]$.
The constructed clause $c^1$ then encodes the opposite of this satisfying
(partial) assignment $\tau$.
We remark that finding such a $\tau$ in the premise of \ref{Rule:solution} is
{\rm NP}-hard (equivalent to finding a solution of a propositional formula in
CNF). This strong condition on $\tau$ is not essential for soundness and
completeness and could be removed\footnote{Then, Lemma~\ref{Lemma:Soundness}
must be weakened to $Pr(Q_{i+1} x_{i+1} \ldots Q_n x_n:
\varphi[\tau(x_1)/x_1]\ldots[\tau(x_{i})/x_{i}]) \leq p$, as for original
S-resolution~\cite{TeiFrae:LPAR17}.} but, as mentioned above, facilitates a less
technical presentation of generalized interpolation in
Section~\ref{Subec:InterpolatingS-Resol}.
Another argument justifying the strong premise of \ref{Rule:solution} is a
potential integration of S-resolution into DPLL-based SSAT solvers since
whenever a satisfying (partial) assignment $\tau$ of $\varphi$ is found by an
SSAT solver then $\tau$ meets the requirements of \ref{Rule:solution}.
\begin{RULE}\label{Rule:solution}
\algrule%
{%
c \subseteq \{x, \neg x| x \in {\it Var}(\varphi)\},
\not\models c, \Q(c)=Q_1 x_1 \ldots Q_i x_i,\\
\text{~for each~} \tau: {\it Var}(\varphi)\downarrow_i \to
\Bool \text{~with~} \forall x \in {\it Var}(c): \tau(x) = \ff_c(x):\\
\models \varphi[\tau(x_1)/x_1]\ldots[\tau(x_{i})/x_{i}]
}%
{%
c^1
}%
\end{RULE}%
Rule~\ref{Rule:resolution} finally constitutes the actual resolution rule as
known from the non-stochastic case. Depending on whether an existential or a
randomized variable is resolved upon, the probability value of the resolvent
clause is computed according to the semantics $Pr(\Q:\varphi)$ defined in
Section~\ref{Sec:Prelim}.
\begin{RULE}\label{Rule:resolution}
\algrule%
{%
(c_1 \vee \neg x)^{p_1}, (c_2 \vee x)^{p_2},
Q x \in \Q, Q x \notin \Q(c_1 \vee c_2),
\not\models (c_1 \vee c_2),\\
p = \left\{
\begin{array}{l@{~;~}l}
\max(p_1,p_2) & Q = \exists\\
p_x \cdot p_1 + (1-p_x) \cdot p_2  & Q = \R^{p_x}\\
\end{array}\right.
}%
{%
(c_1 \vee c_2)^{p}
}%
\end{RULE}%
The derivation of a clause $c^p$
by \ref{Rule:conflict} from $c$,
by \ref{Rule:solution}, and
by \ref{Rule:resolution} from $c_1^{p_1}, c_2^{p_2}$ is denoted
by $c \vdash_{\text{\ref{Rule:conflict}}} c^p$,
by $\vdash_{\text{\ref{Rule:solution}}} c^p$, and
by $(c_1^{p_1}, c_2^{p_2}) \vdash_{\text{\ref{Rule:resolution}}} c^p$,
respectively.
Given rules~\ref{Rule:conflict} to \ref{Rule:resolution}, S-resolution is sound
and complete in the following sense.
\begin{lem}\label{Lemma:Soundness}
Let clause $c^p$ be derivable by S-resolution and let $\Q(c) = Q_1 x_1
\ldots Q_i x_i$. For each $\tau: {\it Var}(\varphi)\downarrow_i \to
\Bool$ with $\forall x \in {\it Var}(c): \tau(x) = \ff_c(x)$ it holds that
$Pr(Q_{i+1} x_{i+1} \ldots Q_n x_n:
\varphi[\tau(x_1)/x_1]\ldots[\tau(x_{i})/x_{i}]) = p.$
\end{lem}
\proof
We show the lemma by induction over the application of
rules~\ref{Rule:conflict}, \ref{Rule:solution}, and
\ref{Rule:resolution}.
The base case is given by rules~\ref{Rule:conflict} and \ref{Rule:solution}.
By construction of $\tau$, $\varphi[\tau(x_1)/x_1]\ldots[\tau(x_{i})/x_{i}]$ is
unsatisfiable for~\ref{Rule:conflict} and tautological for \ref{Rule:solution}
which immediately establishes the result for the base case.
Now assume that the assumption holds for all clauses in the premises
of~\ref{Rule:resolution}, i.e.\
\[
\begin{array}{ll}
Pr(Q_{j+1} x_{j+1} \ldots Q_n x_n:
\varphi[\tau_1(x_1)/x_1]\ldots[\tau_1(x_{j-1})/x_{j-1}][\true/x_j]) & = p_1,\\
Pr(Q_{j+1} x_{j+1} \ldots Q_n x_n:
\varphi[\tau_2(x_1)/x_1]\ldots[\tau_2(x_{j-1})/x_{j-1}][\false/x_j]) & = p_2,\\
\end{array}
\]
where $x_j = x$ with $j \geq i+1$. By definition of $Pr$, for each $\tau$
with $\tau(x)=\tau_1(x)$ if $x \in {\it Var}(c_1)$ and $\tau(x)=\tau_2(x)$ if $x
\in {\it Var}(c_2)$ we then have
\[
\begin{array}{ll}
Pr(Q_j x_j\ Q_{j+1} x_{j+1} \ldots Q_n x_n:
\varphi[\tau(x_1)/x_1]\ldots[\tau(x_{j-1})/x_{j-1}]) & = p.\\
\end{array}
\]
The result is obvious for $j=i+1$. For $j > i+1$, note that variables $x_{i+1},
\ldots, x_{j-1}$ do not occur in the derived clause $(c_1 \vee c_2)$. Hence, for
$k=j-1$ down to $i+1$ we successively conclude that
\[
\begin{array}{ll}
Pr(Q_{k+1} x_{k+1} \ldots Q_n x_n:
\varphi[\tau(x_1)/x_1]\ldots[\tau(x_{k-1})/x_{k-1}][\true/x_{k}])
& = p,\\
Pr(Q_{k+1} x_{k+1} \ldots Q_n x_n:
\varphi[\tau(x_1)/x_1]\ldots[\tau(x_{k-1})/x_{k-1}][\false/x_{k}])
& = p.
\end{array}
\]
From case $k=i+1$ the lemma follows.\qed
\begin{cor}[Soundness of S-resolution]
\label{Theorem:Soundness}
If the empty clause $\emptyset^p$ is derivable by S-resolution from a given SSAT
formula $\Q: \varphi$ then $Pr(\Q: \varphi) = p$.\qed
\end{cor}
Corollary~\ref{Theorem:Soundness} follows directly from
Lemma~\ref{Lemma:Soundness}, namely for the special case $c^p = \emptyset^p$.
Theorem~\ref{Theorem:Completeness} shows completeness of S-resolution.
\begin{thm}[Completeness of S-resolution]
\label{Theorem:Completeness}
If $Pr(\Q: \varphi) = p$ for some SSAT formula $\Q: \varphi$ then the empty
clause $\emptyset^{p}$ is derivable from $\Q: \varphi$ by S-resolution.
\end{thm}
\proof
If $\emptyset \in \varphi$, i.e.\ $\varphi$ contains the empty clause, then
$p=0$ and the empty clause $\emptyset^{0}$ is derivable by
rule~\ref{Rule:conflict}. In the remaining proof, we assume that $\emptyset
\notin \varphi$.
We prove the theorem by induction over the number of quantifiers in the
quantifier prefix $\Q$.
For the base case $\Q=Q x$ we distinguish three cases:
1) $\varphi = (\neg x) \wedge (x)$. Then $p=0$, and $(\neg x)^0$, $(x)^0$ are
derivable by \ref{Rule:conflict}, and \ref{Rule:resolution} finally yields
$\emptyset^{0}$.
2) $\varphi = (\neg x)$. Clauses $(\neg x)^0$ and $(x)^1$ are derivable by
\ref{Rule:conflict} and \ref{Rule:solution}, respectively, the latter since
$\models \varphi[\false/x]$. If $Q=\exists$ or $Q=\R^{p_x}$ then $p=1$ or
$p=(1-p_x)$, and $\emptyset^1$ or $\emptyset^{(1-p_x)}$ can be derived by
\ref{Rule:resolution}, respectively.
3) $\varphi = (x)$. Analogously to 2), if $Q=\exists$ or $Q=\R^{p_x}$ then $p=1$
or $p=p_x$, and $\emptyset^1$ or $\emptyset^{p_x}$ can be derived by
\ref{Rule:resolution}, respectively.

In the induction step, we show that $\emptyset^{p}$ is derivable for
$Pr(Q x\,\Q: \varphi) = p$.
Let $p_1 = Pr(\Q: \varphi[\true/x])$ and $p_2 = Pr(\Q: \varphi[\false/x])$.
Induction hypothesis assumes that $\emptyset^{p_1}$ and $\emptyset^{p_2}$
are derivable from $\Q: \varphi[\true/x]$ and $\Q: \varphi[\false/x]$.
Applying the resolution sequence deriving $\emptyset^{p_1}$ from $\Q:
\varphi[\true/x]$ on $Q x\ \Q: \varphi$ yields either $\emptyset^{p_1}$ or
$(\neg x)^{p_1}$. Analogously, either $\emptyset^{p_2}$ or $(x)^{p_2}$ is
derivable from $Q x\ \Q: \varphi$.
If $\emptyset^{p_1}$ (respectively, $\emptyset^{p_2}$) was derived then $p=p_1$
(respectively, $p=p_2$) by Corollary~\ref{Theorem:Soundness}. (Note that if both
$\emptyset^{p_1}$ and $\emptyset^{p_2}$ are derivable then $p_1=p_2$.)
Otherwise, i.e.\ $(\neg x)^{p_1}$ and $(x)^{p_2}$ are derived, application of
\ref{Rule:resolution} gives $\emptyset^{p}$.\qed

The above presentation of S-resolution differs slightly from
\cite{TeiFrae:LPAR17} in order to avoid overhead in interpolant generation
incurred when employing the original definition, like the necessity of enforcing
particular resolution sequences. For readers familiar with
\cite{TeiFrae:LPAR17}, the particular modifications are:
1) derived clauses $c^p$ may also carry value $p=1$,
2) former rules~R.2 and R.5 are joined into the new rule~\ref{Rule:solution},
and
3) former rules~R.3 and R.4 are collapsed into rule~\ref{Rule:resolution}.
These modifications do not affect soundness and completeness of S-resolution,
confer Corollary~\ref{Theorem:Soundness} and
Theorem~\ref{Theorem:Completeness}. The advantage of the modification is that
derivable clauses $c^p$ are forced to have a tight bound~$p$ in the sense that
under each assignment which falsifies $c$, the satisfaction probability of the
remaining subproblem \emph{exactly} is $p$, confer Lemma~\ref{Lemma:Soundness}.
This fact confirms the conjecture from~\cite[page~14]{TeiFrae:LPAR17} about the
existence of such clauses $(c \vee \ell)^p$ and allows for a generalized clause
learning scheme to be integrated into DPLL-SSAT solvers: the idea is that under
a partial assignment falsifying $c$, one may directly propagate literal~$\ell$
as the satisfaction probability of the other branch, for which the negation of
$\ell$ holds, is known to be $p$ already.

\paragraph{Example of S-resolution.}
Consider the SSAT formula $\Phi = \R^{0.8}x_1\ \exists x_2\ \R^{0.3}x_3: ((x_1
\vee x_2) \wedge (\neg x_2) \wedge (x_2 \vee x_3))$ with $Pr(\Phi)=0.24$.
Clauses $(x_1 \vee x_2)^0$, $(\neg x_2)^0$, $(x_2 \vee x_3)^0$ are then
derivable by \ref{Rule:conflict}. As $x_1=\true, x_2=\false, x_3=\true$ is a
satisfying assignment, $\vdash_{\text{\ref{Rule:solution}}}
(\neg x_1 \vee x_2 \vee \neg x_3)^1$.
Then, $((\neg x_1 \vee x_2 \vee \neg x_3)^1, (x_2 \vee x_3)^0)
\vdash_{\text{\ref{Rule:resolution}}} (\neg x_1 \vee x_2)^{0.3}$,
$((\neg x_2)^0, (\neg x_1 \vee x_2)^{0.3})
\vdash_{\text{\ref{Rule:resolution}}} (\neg x_1)^{0.3}$,
$((\neg x_2)^0, (x_1 \vee x_2)^0)
\vdash_{\text{\ref{Rule:resolution}}} (x_1)^{0}$, and finally
$((\neg x_1)^{0.3}, (x_1)^{0})
\vdash_{\text{\ref{Rule:resolution}}} \emptyset^{0.24}$.

\subsection{Interpolating resolution for SSAT}
\label{Subec:InterpolatingS-Resol}
We now devote our attention to the computation of generalized Craig
interpolants for SSAT by means of an enhanced version of S-resolution, which is
akin to resolution-based Craig interpolation for propositional SAT
by Pudl\'{a}k~\cite{Pudlak97}. We remark that on SSAT formulae $\Q: (A
\wedge B)$, Pudl\'{a}k's algorithm, which has unsatisfiability of $A \wedge B$
as precondition, will not work in general. When instead considering
the unsatisfiable formula $A \wedge B \wedge \neg \sol$ with $\neg
\sol$ in CNF then Pudl\'{a}k's method would be applicable and would
actually produce a generalized Craig interpolant. The main drawback of this
approach however is the explicit construction of $\neg \sol$, calling for
explicit quantifier elimination.

In the following, we propose an algorithm based on S-resolution for computing
generalized Craig interpolants which operates directly on $A \wedge B$ without
adding $\neg \sol$, and thus does not comprise any preprocessing involving
quantifier elimination. For this purpose, the rules of S-resolution are enhanced
to deal with pairs $(c^p, I)$ of annotated clauses $c^p$ and propositional
formulae $I$. Such formulae $I$ are in a certain sense \emph{intermediate}
generalized interpolants, i.e.\ generalized interpolants for subformulae
arising from instantiating some variables by partial assignments that
falsify $c$, confer Lemma~\ref{Lemma:InterpolantsComputation}. Once a
pair $(\emptyset^p, I)$ comprising the empty clause is derived, $I$
thus is a generalized Craig interpolant for the given SSAT
formula. This augmented S-resolution, which we call
\emph{interpolating S-resolution}, is defined by
rules~\ref{Rule2:conflict}, \ref{Rule2:solution}, and
\ref{Rule2:resolution}. The construction of intermediate interpolants
$I$ in \ref{Rule2:conflict} and \ref{Rule2:resolution} coincides with
the classical rules by Pudl\'{a}k~\cite{Pudlak97}, while
\ref{Rule2:solution} misses a corresponding counterpart. The rationale
is that \ref{Rule2:solution} (or rather \ref{Rule:solution}) refers to
satisfying valuations $\tau$ of $A \wedge B$, which do not exist in
classical interpolation.  As $A \wedge B$ becomes a tautology after
substituting the partial assignment $\tau$ from \ref{Rule:solution}
into it, its quantified variant $\sol=\exists
a_1,\ldots,b_1,\ldots:A\wedge B$ also becomes tautological under the
same substitution
$\sol[\tau(x_1)/x_1,\ldots,\tau(x_i)/x_i]$. Consequently,
$\neg\sol[\tau(x_1)/x_1,\ldots,\tau(x_i)/x_i]$ is unsatisfiable, and
so are $(A \wedge \neg \sol)[\tau(x_1)/x_1,\ldots,\tau(x_i)/x_i]$ and
$(B \wedge \neg \sol)[\tau(x_1)/x_1,\ldots,\tau(x_i)/x_i]$. This
implies that the actual intermediate interpolant in
\ref{Rule2:solution} can be chosen arbitrarily over variables in
$V_{A,B}$. This freedom will allow us to control the geometric extent
of generalized interpolants within the ``don't care''-region provided by
the models of $\sol$, confer Corollary~\ref{Cor:ControlSIntComp}.
\begin{RULE2}\label{Rule2:conflict}
\algrule%
{%
c \vdash_{\text{\ref{Rule:conflict}}} c^p,
I = \left\{
\begin{array}{l@{~;~}l}
\false & c \in A\\
\true  & c \in B\\
\end{array}\right.
}%
{%
(c^p, I)
}%
\end{RULE2}%
\begin{RULE2}\label{Rule2:solution}
\algrule%
{%
\vdash_{\text{\ref{Rule:solution}}} c^p,
I \text{~is~any~formula~over~} V_{A,B}
}%
{%
(c^{p}, I)
}%
\end{RULE2}%
\begin{RULE2}\label{Rule2:resolution}
\algrule%
{%
((c_1 \vee \neg x)^{p_1}, I_1),
((c_2 \vee x)^{p_2}, I_2),\\
((c_1 \vee \neg x)^{p_1}, (c_2 \vee x)^{p_2})
\vdash_{\text{\ref{Rule:resolution}}} (c_1 \vee c_2)^p,\\
I = \left\{
\begin{array}{c@{~;~}l}
I_1 \vee I_2 & x \in V_A\\
I_1 \wedge I_2 & x \in V_B\\
(\neg x \vee I_1) \wedge (x \vee I_2) & x \in V_{A,B}\\
\end{array}\right.\\
}%
{%
((c_1 \vee c_2)^p, I)
}%
\end{RULE2}%
The following lemma establishes the theoretical foundation of
computing generalized Craig interpolants by interpreting the derived pairs
$(c^p, I)$.
\begin{lem}\label{Lemma:InterpolantsComputation}
Let $\Phi = \Q: (A \wedge B)$ with $\Q=Q_1 x_1 \ldots Q_n x_n$ be some SSAT
formula, and the pair $(c^p, I)$ be derivable from $\Phi$ by interpolating
S-resolution, where $\Q(c) = Q_1 x_1 \ldots Q_i x_i$.
Then, for each $\tau: {\it Var}(A \wedge B)\downarrow_i \to
\Bool$ with $\forall x \in {\it Var}(c): \tau(x) = \ff_c(x)$ it
holds that
\begin{enumerate}[\em(1)]
\item\label{item:globalVarsI}
${\it Var}(I) \subseteq V_{A,B}$,
\item\label{item:IntA}
$Pr(Q_{i+1} x_{i+1} \ldots Q_n x_n: (A \wedge \neg \sol \wedge \neg
I)[\tau(x_1)/x_1]\ldots[\tau(x_{i})/x_{i}]) = 0$, and
\item\label{item:IntB}
$Pr(Q_{i+1} x_{i+1} \ldots Q_n x_n: (I \wedge B \wedge \neg
\sol)[\tau(x_1)/x_1]\ldots[\tau(x_{i})/x_{i}])
= 0$.
\end{enumerate}
\end{lem}
\proof
We prove the lemma by induction over application of the interpolating
S-resolution rules~\ref{Rule2:conflict}, \ref{Rule2:solution}, and
\ref{Rule2:resolution}.
In the base case, we can just apply \ref{Rule2:conflict} and
\ref{Rule2:solution}. Item~\ref{item:globalVarsI} clearly holds for both
rules since $I$ contains only variables in $V_{A,B}$.
Let us consider \ref{Rule2:conflict} first.
If $c \in A$ then $I=\false$. By construction of $\tau$, i.e.\ $c$ evaluates to
$\false$ under $\tau$, it follows that
$A[\tau(x_1)/x_1]\ldots[\tau(x_{i})/x_{i}]$ is unsatisfiable and thus
\[
Pr(\Q': (A \wedge \neg \sol \wedge \neg
I)[\tau(x_1)/x_1]\ldots[\tau(x_{i})/x_{i}]) = 0~~.
\]
As $I=\false$, immediately
\[
Pr(\Q': (I \wedge B \wedge \neg \sol)[\tau(x_1)/x_1]\ldots[\tau(x_{i})/x_{i}])
= 0~~.
\]
If $c \in B$ then $I=\true$. Obviously,
\[
Pr(\Q': (A \wedge \neg \sol \wedge \neg
I)[\tau(x_1)/x_1]\ldots[\tau(x_{i})/x_{i}]) = 0
\]
and by construction of $\tau$,
\[
Pr(\Q': (I \wedge B \wedge \neg \sol)[\tau(x_1)/x_1]\ldots[\tau(x_{i})/x_{i}])
= 0~~.
\]
For rule~\ref{Rule2:solution}, we have $\models (A \wedge
B)[\tau(x_1)/x_1]\ldots[\tau(x_{i})/x_{i}]$ which immediately implies
that $\models (\exists a_1, \ldots, a_{\alpha}, b_1, \ldots, b_{\beta}: (A
\wedge B))[\tau(x_1)/x_1]\ldots[\tau(x_{i})/x_{i}]$, i.e.\
$\models \sol[\tau(x_1)/x_1]\ldots[\tau(x_{i})/x_{i}]$ by definition of
$\sol$. Rephrasing the latter,
$\neg \sol[\tau(x_1)/x_1]\ldots[\tau(x_{i})/x_{i}]$ is unsatisfiable.
Consequently, for any propositional formula $I$
\begin{align*}
Pr(\Q': (A \wedge \neg \sol \wedge \neg
I)[\tau(x_1)/x_1]\ldots[\tau(x_{i})/x_{i}]) & = 0~~,\\
Pr(\Q': (I \wedge B \wedge \neg \sol)[\tau(x_1)/x_1]\ldots[\tau(x_{i})/x_{i}])
& = 0~~.
\end{align*}
This proves items~\ref{item:IntA} and \ref{item:IntB} for the base case.

In the induction step, we now assume that the lemma holds for all clauses in the
premises of rule~\ref{Rule2:resolution}. Then, by construction of $I$,
item~\ref{item:globalVarsI} clearly holds for $I$, i.e.\ ${\it Var}(I) \subseteq
V_{A,B}$. Induction hypothesis assumes that
\begin{align*}
Pr(\Q': (A \wedge \neg \sol \wedge \neg
I_1)[\tau_1(x_1)/x_1]\ldots[\tau_1(x_{j-1})/x_{j-1}][\true/x_j]) & = 0~~,\\
Pr(\Q': (I_1 \wedge B \wedge \neg
\sol)[\tau_1(x_1)/x_1]\ldots[\tau_1(x_{j-1})/x_{j-1}][\true/x_j]) & = 0
\end{align*}
holds for $((c_1 \vee \neg x_j)^{p_1}, I_1)$ and for each
$\tau_1: {\it Var}(A \wedge B)\downarrow_{j-1} \to \Bool$ with $\forall x \in
{\it Var}(c_1): \tau_1(x) = \ff_{c_1}(x)$, and that
\begin{align*}
Pr(\Q': (A \wedge \neg \sol \wedge \neg
I_2)[\tau_2(x_1)/x_1]\ldots[\tau_2(x_{j-1})/x_{j-1}][\false/x_j]) & = 0~~,\\
Pr(\Q': (I_2 \wedge B \wedge \neg
\sol)[\tau_2(x_1)/x_1]\ldots[\tau_2(x_{j-1})/x_{j-1}][\false/x_j]) & = 0
\end{align*}
holds for $((c_2 \vee x_j)^{p_2}, I_2)$ and for each $\tau_2: {\it Var}(A \wedge
B)\downarrow_{j-1} \to \Bool$ with $\forall x \in {\it Var}(c_2): \tau_2(x) =
\ff_{c_2}(x)$,
where $j \geq i+1$ and $\Q' = Q_{j+1} x_{j+1} \ldots Q_{n} x_{n}$.
Let $\tau: {\it Var}(A \wedge B)\downarrow_{j-1} \to \Bool$ be any assignment
with $\tau(x)=\tau_1(x)$ if $x \in {\it Var}(c_1)$ and
$\tau(x)=\tau_2(x)$ if $x \in {\it Var}(c_2)$. Note that $\tau$ is well-defined
as $\not\models (c_1 \vee c_2)$, i.e.\ for each $x \in {\it Var}(c_1) \cap {\it
Var}(c_2): \tau_1(x) = \tau_2(x)$.
We now show that
\begin{align*}
Pr_A & :=
Pr(Q_j x_j \Q': (A \wedge \neg \sol \wedge \neg
I)[\tau(x_1)/x_1]\ldots[\tau(x_{j-1})/x_{j-1}]) & & = 0~~,\\
Pr_B & :=
Pr(Q_j x_j \Q': (I \wedge B \wedge \neg
\sol)[\tau(x_1)/x_1]\ldots[\tau(x_{j-1})/x_{j-1}]) & & = 0
\end{align*}
by proving that
\begin{align*}
Pr_{A,x} & :=
Pr(\Q': (A \wedge \neg \sol \wedge \neg
I)[\tau(x_1)/x_1]\ldots[\tau(x_{j-1})/x_{j-1}][\true/x_j]) && = 0~~,\\
Pr_{A,\neg x} & :=
Pr(\Q': (A \wedge \neg \sol \wedge \neg
I)[\tau(x_1)/x_1]\ldots[\tau(x_{j-1})/x_{j-1}][\false/x_j]) && = 0~~,\\
Pr_{B,x} & :=
Pr(\Q': (I \wedge B \wedge \neg
\sol)[\tau(x_1)/x_1]\ldots[\tau(x_{j-1})/x_{j-1}][\true/x_j]) && = 0~~,\\
Pr_{B,\neg x} & :=
Pr(\Q': (I \wedge B \wedge \neg
\sol)[\tau(x_1)/x_1]\ldots[\tau(x_{j-1})/x_{j-1}][\false/x_j]) && = 0~~.
\end{align*}
We therefore distinguish the three cases $x_j \in V_A$, $x_j \in V_B$, and $x_j
\in V_{A,B}$.

First, let be $x_j \in V_A$. Then, $I= I_1 \vee I_2$.
By induction hypothesis and by construction of $I$,
\begin{align*}
0 & = Pr(\Q': (A \wedge \neg \sol \wedge \neg
I_1)[\tau_1(x_1)/x_1]\ldots[\tau_1(x_{j-1})/x_{j-1}][\true/x_j])\\
  & \geq Pr(\Q': (A \wedge \neg \sol \wedge \neg I_1 \wedge \neg
I_2)[\tau_1(x_1)/x_1]\ldots[\tau_1(x_{j-1})/x_{j-1}][\true/x_j])\\
  & = Pr(\Q': (A \wedge \neg \sol \wedge \neg
I)[\tau_1(x_1)/x_1]\ldots[\tau_1(x_{j-1})/x_{j-1}][\true/x_j])~~.\\
\intertext{Due to construction of $\tau$, it holds in particular that}
0 & = Pr_{A,x}~~.\\
\intertext{Analogously,}
0 & = Pr(\Q': (A \wedge \neg \sol \wedge \neg
I_2)[\tau_2(x_1)/x_1]\ldots[\tau_2(x_{j-1})/x_{j-1}][\false/x_j])\\
  & \geq Pr(\Q': (A \wedge \neg \sol \wedge \neg I_1 \wedge \neg
I_2)[\tau_2(x_1)/x_1]\ldots[\tau_2(x_{j-1})/x_{j-1}][\false/x_j])\\
  & = Pr(\Q': (A \wedge \neg \sol \wedge \neg
I)[\tau_2(x_1)/x_1]\ldots[\tau_2(x_{j-1})/x_{j-1}][\false/x_j])\\
\intertext{and thus}
0 & = Pr_{A,\neg x}~~.
\end{align*}
As $x_j \notin {\it Var}(I) \cup {\it Var}(B) \cup {\it Var}(\neg \sol)$,
for each $v \in \Bool$ it holds that
\begin{align*}
& Pr(\Q': (I \wedge B \wedge \neg
\sol)[\tau(x_1)/x_1]\ldots[\tau(x_{j-1})/x_{j-1}][v/x_j])\\
& = Pr(\Q': (I
\wedge B \wedge \neg
\sol)[\tau(x_1)/x_1]\ldots[\tau(x_{j-1})/x_{j-1}])
\end{align*}
which implies $Pr_{B,x} = Pr_{B,\neg x}$.
We conclude from induction hypothesis that
\begin{align*}
Pr(\Q': (I_1 \wedge B \wedge \neg
\sol)[\tau_1(x_1)/x_1]\ldots[\tau_1(x_{j-1})/x_{j-1}]) & = 0~~,\\
Pr(\Q': (I_2 \wedge B \wedge \neg
\sol)[\tau_2(x_1)/x_1]\ldots[\tau_2(x_{j-1})/x_{j-1}]) & = 0
\end{align*}
again by virtue of $x_j \notin {\it Var}(I) \cup {\it Var}(B) \cup {\it
Var}(\neg \sol)$.
Moreover,
\begin{align*}
Pr(\Q': (I_1 \wedge B \wedge \neg
\sol)[\tau(x_1)/x_1]\ldots[\tau(x_{j-1})/x_{j-1}]) & = 0~~,\\
Pr(\Q': (I_2 \wedge B \wedge \neg
\sol)[\tau(x_1)/x_1]\ldots[\tau(x_{j-1})/x_{j-1}]) & = 0
\end{align*}
due to construction of $\tau$.
Note that if $Pr(\Q:\varphi_1) = 0$ and $Pr(\Q:\varphi_2) = 0$ then
$Pr(\Q:(\varphi_1 \vee \varphi_2)) = 0$ since $Pr(\Q:\varphi) = 0$ if and only
if $\varphi$ is unsatisfiable.\footnote{This statement is not true in general if
$\Q$ also contains \emph{universal} quantifiers, which is not the case in this
article. However, extensions of SSAT involving universal quantifiers have also
been considered in the literature, confer~\cite{Maj09HBSAT}.} As a consequence,
\begin{align*}
0 & =
Pr\left(\Q': \left(
\begin{array}{ll}
& (I_1 \wedge B \wedge \neg
\sol)[\tau(x_1)/x_1]\ldots[\tau(x_{j-1})/x_{j-1}]\\
\vee & (I_2 \wedge B \wedge \neg
\sol)[\tau(x_1)/x_1]\ldots[\tau(x_{j-1})/x_{j-1}]\\
\end{array}\right)\right)\\
  & =
Pr(\Q': ((I_1 \vee I_2) \wedge B \wedge \neg
\sol)[\tau(x_1)/x_1]\ldots[\tau(x_{j-1})/x_{j-1}])\\
  & =
Pr(\Q': (I \wedge B \wedge \neg
\sol)[\tau(x_1)/x_1]\ldots[\tau(x_{j-1})/x_{j-1}])\\
  & = Pr_{B,x} = Pr_{B,\neg x}~~.
\end{align*}

Second, let be $x_j \in V_B$. Then, $I= I_1 \wedge I_2$.
As $x_j \notin {\it Var}(A) \cup {\it Var}(\neg \sol) \cup {\it Var}(\neg I)$,
with the same argument as above,
\begin{align*}
0 & =
Pr\left(\Q': \left(
\begin{array}{ll}
& (A \wedge \neg \sol \wedge \neg
I_1)[\tau(x_1)/x_1]\ldots[\tau(x_{j-1})/x_{j-1}]\\
\vee & (A \wedge \neg \sol \wedge \neg
I_2)[\tau(x_1)/x_1]\ldots[\tau(x_{j-1})/x_{j-1}]\\
\end{array}\right)\right)\\
  & =
Pr(\Q': (A \wedge \neg \sol \wedge
(\neg I_1 \vee \neg I_2))[\tau(x_1)/x_1]\ldots[\tau(x_{j-1})/x_{j-1}])\\
  & =
Pr(\Q': (A \wedge \neg \sol \wedge \neg
I)[\tau(x_1)/x_1]\ldots[\tau(x_{j-1})/x_{j-1}])\\
  & = Pr_{A,x} = Pr_{A,\neg x}~~.
\end{align*}
Again following the reasoning above, we have
\begin{align*}
0 & = Pr(\Q': (I_1 \wedge B \wedge \neg
\sol)[\tau_1(x_1)/x_1]\ldots[\tau_1(x_{j-1})/x_{j-1}][\true/x_j])\\
  & \geq Pr(\Q': (I_1 \wedge I_2 \wedge B \wedge \neg
\sol)[\tau_1(x_1)/x_1]\ldots[\tau_1(x_{j-1})/x_{j-1}][\true/x_j])\\
  & = Pr(\Q': (I \wedge B \wedge \neg
\sol)[\tau_1(x_1)/x_1]\ldots[\tau_1(x_{j-1})/x_{j-1}][\true/x_j])\\
\intertext{and thus}
0 & = Pr_{B,x}\\
\intertext{as well as}
0 & = Pr(\Q': (I_2 \wedge B \wedge \neg
\sol)[\tau_2(x_1)/x_1]\ldots[\tau_2(x_{j-1})/x_{j-1}][\false/x_j])\\
  & \geq Pr(\Q': (I_1 \wedge I_2 \wedge B \wedge \neg
\sol)[\tau_2(x_1)/x_1]\ldots[\tau_2(x_{j-1})/x_{j-1}][\false/x_j])\\
  & = Pr(\Q': (I \wedge B \wedge \neg
\sol)[\tau_2(x_1)/x_1]\ldots[\tau_2(x_{j-1})/x_{j-1}][\false/x_j])~~,\\
\intertext{and thus}
0 & = Pr_{B,\neg x}~~.
\end{align*}

Third, let be $x_j \in V_{A,B}$. Then, $I= (\neg x_j \vee I_1) \wedge (x_j
\vee I_2)$, and we deduce
\begin{align*}
0 & =
Pr(\Q': (A \wedge \neg \sol \wedge \neg
I_1)[\tau_1(x_1)/x_1]\ldots[\tau_1(x_{j-1})/x_{j-1}][\true/x_j])\\
  & =
Pr(\Q': (A \wedge \neg \sol\\
&\quad\quad\quad\quad \wedge ((x_j \wedge \neg I_1) \vee (\neg x_j
\wedge \neg I_2)))[\tau_1(x_1)/x_1]\ldots[\tau_1(x_{j-1})/x_{j-1}][\true/x_j])\\
  & =
Pr(\Q': (A \wedge \neg \sol \wedge \neg
I)[\tau_1(x_1)/x_1]\ldots[\tau_1(x_{j-1})/x_{j-1}][\true/x_j])\\
\intertext{and, in particular,}
0 & =
Pr_{A,x}~~.\\
\intertext{Analogously,}
0 & =
Pr(\Q': (A \wedge \neg \sol \wedge \neg
I_2)[\tau_2(x_1)/x_1]\ldots[\tau_2(x_{j-1})/x_{j-1}][\false/x_j])\\
  & =
Pr(\Q': (A \wedge \neg \sol\\
&\quad\quad\quad\quad \wedge ((x_j \wedge \neg I_1) \vee (\neg x_j \wedge \neg
I_2)))[\tau_2(x_1)/x_1]\ldots[\tau_2(x_{j-1})/x_{j-1}][\false/x_j])\\
  & =
Pr(\Q': (A \wedge \neg \sol \wedge \neg
I)[\tau_2(x_1)/x_1]\ldots[\tau_2(x_{j-1})/x_{j-1}][\false/x_j])\\
\intertext{and, in particular,}
0 & =
Pr_{A,\neg x}~~.\\
\intertext{Furthermore,}
0 & =
Pr(\Q': (I_1 \wedge B \wedge \neg
\sol)[\tau_1(x_1)/x_1]\ldots[\tau_1(x_{j-1})/x_{j-1}][\true/x_j])\\
  & =
Pr(\Q': ((\neg x_j \vee I_1) \wedge (x_j \vee I_2) \wedge B\\
&\quad\quad\quad\quad \wedge \neg
\sol)[\tau_1(x_1)/x_1]\ldots[\tau_1(x_{j-1})/x_{j-1}][\true/x_j])\\
  & =
Pr(\Q': (I \wedge B \wedge \neg
\sol)[\tau_1(x_1)/x_1]\ldots[\tau_1(x_{j-1})/x_{j-1}][\true/x_j])\\
\intertext{and, in particular,}
0 & =
Pr_{B,x}~~.\\
\intertext{Finally,}
0 & =
Pr(\Q': (I_2 \wedge B \wedge \neg
\sol)[\tau_2(x_1)/x_1]\ldots[\tau_2(x_{j-1})/x_{j-1}][\false/x_j])\\
  & =
Pr(\Q': ((\neg x_j \vee I_1) \wedge (x_j \vee I_2) \wedge B\\
&\quad\quad\quad\quad \wedge \neg
\sol)[\tau_2(x_1)/x_1]\ldots[\tau_2(x_{j-1})/x_{j-1}][\false/x_j])\\
  & =
Pr(\Q': (I \wedge B \wedge \neg
\sol)[\tau_2(x_1)/x_1]\ldots[\tau_2(x_{j-1})/x_{j-1}][\false/x_j])\\
\intertext{and, in particular,}
0 & =
Pr_{B,\neg x}~~.\\
\end{align*}
Having shown that $Pr_{A,x} = Pr_{A,\neg x} = Pr_{B,x} = Pr_{B,\neg x} = 0$, we
can now prove the intermediate result above, i.e.\ $Pr_A = Pr_B = 0$. If
$Q_j=\exists$ then $Pr_A = \max(Pr_{A,x},Pr_{A,\neg x}) = 0$ and $Pr_B =
\max(Pr_{B,x},Pr_{B,\neg x}) = 0$, and if $Q_j=\R^{p_x}$ then $Pr_A =
p_x \cdot Pr_{A,x} + (1-p_x) \cdot Pr_{A,\neg x} = 0$ and $Pr_B = p_x \cdot
Pr_{B,x} + (1-p_x) \cdot Pr_{B,\neg x} = 0$.

To finish the proof, we finally need to show that items~\ref{item:IntA} and
\ref{item:IntB}, i.e.\
\begin{align*}
& Pr(Q_{i+1} x_{i+1} \ldots Q_n x_n: (A \wedge \neg \sol \wedge \neg
I)[\tau(x_1)/x_1]\ldots[\tau(x_{i})/x_{i}]) && = 0~~,\\
& Pr(Q_{i+1} x_{i+1} \ldots Q_n x_n: (I \wedge B \wedge \neg
\sol)[\tau(x_1)/x_1]\ldots[\tau(x_{i})/x_{i}]) && = 0~~,
\end{align*}
follow from $Pr_A = Pr_B = 0$, i.e.\ from
\begin{align*}
& Pr(Q_j x_j \ldots Q_n x_n: (A \wedge \neg \sol \wedge \neg
I)[\tau(x_1)/x_1]\ldots[\tau(x_{j-1})/x_{j-1}]) && = 0~~,\\
& Pr(Q_j x_j \ldots Q_n x_n: (I \wedge B \wedge \neg
\sol)[\tau(x_1)/x_1]\ldots[\tau(x_{j-1})/x_{j-1}]) && = 0~~.
\end{align*}
If $j=i+1$ then the result is obvious. Otherwise, i.e.\ if $j>i+1$, the
variables $x_{i+1}, \ldots, x_{j-1}$ do not occur in the derived clause $(c_1
\vee c_2)$ since $\Q(c_1 \vee c_2) = Q_1 x_1 \ldots Q_i x_i$. By definition of
assignment $\tau$, for $k = j-1$ down to $i+1$ we may therefore successively
conclude that
\begin{align*}
& Pr(Q_{k+1} x_{k+1} \ldots Q_n x_n:\\
&\quad\quad\quad\quad\quad\quad (A \wedge \neg \sol \wedge \neg
I)[\tau(x_1)/x_1]\ldots[\tau(x_{k-1})/x_{k-1}][\true/x_k]) && = 0~~,\\
& Pr(Q_{k+1} x_{k+1} \ldots Q_n x_n:\\
&\quad\quad\quad\quad\quad\quad (A \wedge \neg \sol \wedge \neg
I)[\tau(x_1)/x_1]\ldots[\tau(x_{k-1})/x_{k-1}][\false/x_k]) && = 0~~,\\
& Pr(Q_{k+1} x_{k+1} \ldots Q_n x_n:\\
&\quad\quad\quad\quad\quad\quad (I \wedge B \wedge \neg
\sol)[\tau(x_1)/x_1]\ldots[\tau(x_{k-1})/x_{k-1}][\true/x_k]) && = 0~~,\\
& Pr(Q_{k+1} x_{k+1} \ldots Q_n x_n:\\
&\quad\quad\quad\quad\quad\quad (I \wedge B \wedge \neg
\sol)[\tau(x_1)/x_1]\ldots[\tau(x_{k-1})/x_{k-1}][\false/x_k]) && = 0~~.
\end{align*}
From case $k=i+1$ the result immediately follows.\qed
Completeness of S-resolution, as stated in
Theorem~\ref{Theorem:Completeness}, together with above
Lemma~\ref{Lemma:InterpolantsComputation}, applied to the derived pair
$(\emptyset^p, I)$, yields
\begin{cor}[Generalized Craig interpolants computation]
\label{Cor:SIntComp}
If $\Q: (A \wedge B)$ is an SSAT formula then a generalized Craig interpolant
for $(A,B)$ can be computed by interpolating S-resolution.\qed
\end{cor}
Note that computation of generalized interpolants does not depend on
the actual truth state of $A \wedge B$.
The next observation facilitates to effectively control the geometric extent of
generalized Craig interpolants within the ``don't care''-region $\sol$. This
result will be useful within applications of generalized Craig interpolation to
the symbolic analysis of probabilistic systems being investigated in
Section~\ref{Sec:Analysis}.
\begin{cor}[Controlling generalized Craig interpolants computation]
\label{Cor:ControlSIntComp}
If $I=\true$ is used within each application of rule~\ref{Rule2:solution}
then $Pr(\Q: (A \wedge \neg \Int)) = 0$.
Likewise, if $I=\false$ is used in rule~\ref{Rule2:solution} then
$Pr(\Q: (\Int \wedge B)) = 0$.
\end{cor}
\proof
The proof works analogously to the one of
Lemma~\ref{Lemma:InterpolantsComputation}. For the base case, it is clear that
the desired property for \ref{Rule2:conflict} is independent of $\neg \sol$.
For \ref{Rule2:solution}, if $I=\true$ then clearly
$Pr(\Q': (A \wedge \neg I)[\tau(x_1)/x_1]\ldots[\tau(x_{i})/x_{i}])  = 0$, and
if $I=\false$ then
$Pr(\Q': (I \wedge B)[\tau(x_1)/x_1]\ldots[\tau(x_{i})/x_{i}]) = 0$. Then, we
can modify the induction hypothesis: for case ``$I=\true$ in
\ref{Rule2:solution}'', we assume that
$Pr(\Q': (A \wedge \neg I_1)[\tau_1(x_1)/x_1]\ldots[\true/x_j]) = 0$,
$Pr(\Q': (A \wedge \neg I_2)[\tau_2(x_1)/x_1]\ldots[\false/x_j]) = 0$,
and for ``$I=\false$ in \ref{Rule2:solution}'' that
$Pr(\Q': (I_1 \wedge B)[\tau_1(x_1)/x_1]\ldots[\true/x_j]) = 0$,
$Pr(\Q': (I_2 \wedge B)[\tau_2(x_1)/x_1]\ldots[\false/x_j]) = 0$. The induction
step then follows the same reasoning as in the remaining proof of
Lemma~\ref{Lemma:InterpolantsComputation}.\qed
Observe that the special interpolants $\Int$ from
Corollary~\ref{Cor:ControlSIntComp} relate to the classical strongest
and weakest Craig interpolants $A^{\exists}$ and
$\overline{B}^{\forall}$, respectively, in the following sense: $Pr(\Q: (A
\wedge \neg \Int)) = 0$ iff $\models A \implies \Int$ iff $\models
\forall a_1, \ldots, a_{\alpha}: (A \implies \Int)$ iff $\models
(A^{\exists} \implies \Int)$, as $a_1, \ldots, a_{\alpha}$ do not
occur in $\Int$. Analogously, $Pr(\Q: (\Int \wedge B)) = 0$ iff
$\models \Int \implies \neg B$ iff $\models \forall b_1, \ldots,
b_{\beta}: (\Int \implies \neg B)$ iff $\models \Int \implies 
\overline{B}^{\forall}$.

\paragraph{Example of computing generalized Craig interpolants by interpolating
S-resolution}

\begin{figure}[t]
\centering
\resizebox{1.0\textwidth}{!}
{\input{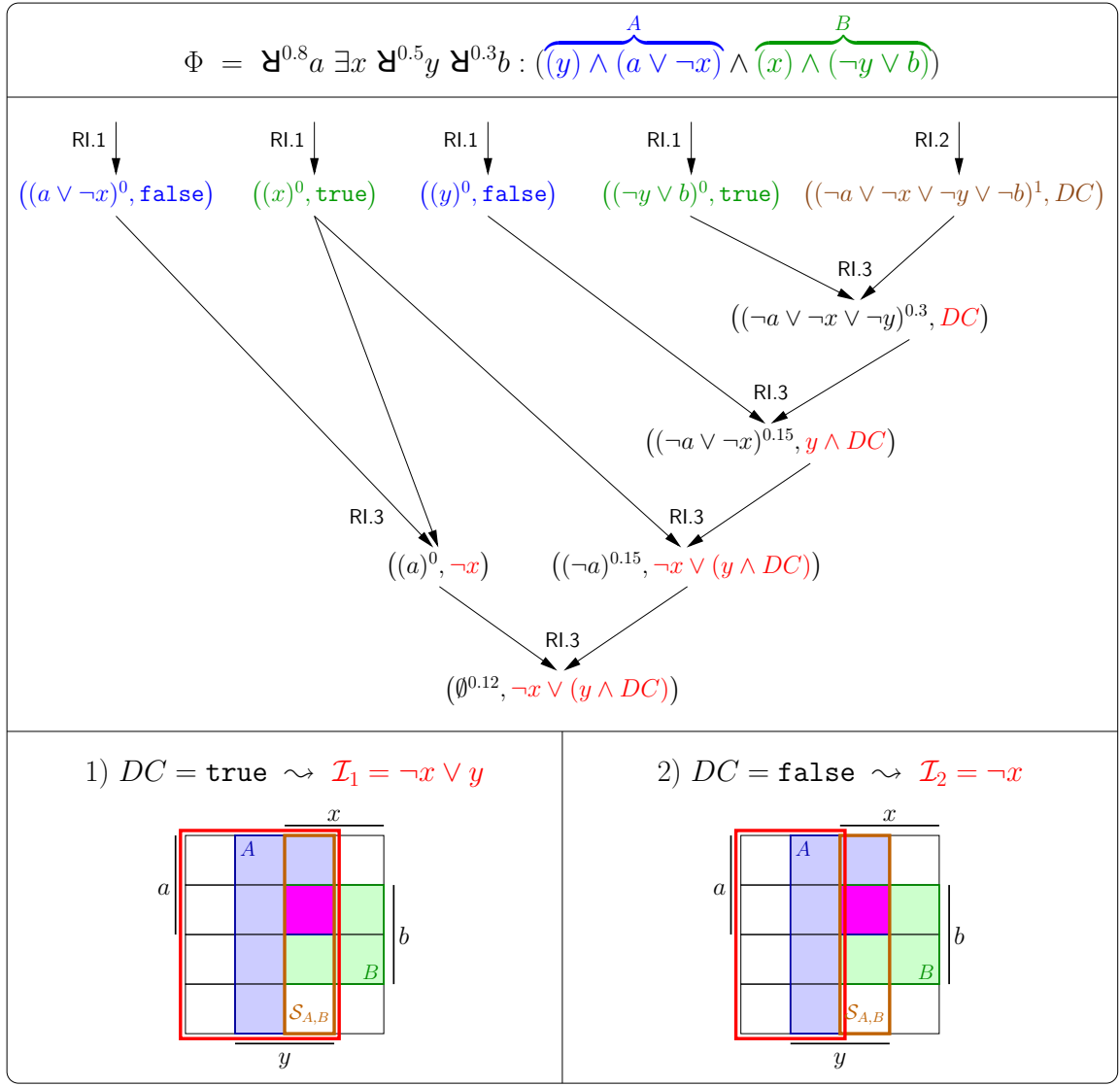}}
\caption{Example of interpolating S-resolution and illustration of the
resulting generalized Craig interpolants by means of Karnaugh-Veitch diagrams.
Arrows denote applications of the specified interpolating S-resolution rules,
while $DC$ stands for any formula over $V_{A,B}$ as in
rule~\ref{Rule2:solution}.}
\label{Fig:ExampleInterpolSres}
\end{figure}

For an example of interpolating S-resolution, consider the SSAT formula
$\Phi = \R^{0.8}a\ \exists x\ \R^{0.5}y\ \R^{0.3}b:
(A \wedge B)$
with $A = ((y) \wedge (a \vee \neg x))$ and $B = ((x) \wedge (\neg y \vee b))$.
Then, $V_A = \{a\}$, $V_B = \{b\}$, and $V_{A,B} = \{x,y\}$.
It is not hard to see that the only satisfying assignment $\tau$ of the
propositional formula $A \wedge B$ is given by $\tau(a)=\true$, $\tau(x)=\true$,
$\tau(y)=\true$, and $\tau(b)=\true$. Hence, $Pr(\Phi) = 0.12$.
A derivation of the empty clause $\emptyset^{0.12}$ together with its associated
generalized Craig interpolant $\neg x \vee (y \wedge DC)$ is shown in
Figure~\ref{Fig:ExampleInterpolSres}, while $DC$ stands for any formula over
variables in $V_{A,B}$ as in rule~\ref{Rule2:solution}.
Note that pair $((\neg a \vee \neg x \vee \neg y \vee \neg b)^1, DC)$ is
derivable by rule~\ref{Rule2:solution} since $\models (A \wedge
B)[\tau(a)/a][\tau(x)/x][\tau(y)/y][\tau(b)/b]$.
Applying Corollary~\ref{Cor:ControlSIntComp} by choosing $DC=\true$ and
$DC=\false$, we obtain the generalized Craig interpolants $\Int_1 = \neg x \vee
y$ and $\Int_2 = \neg x$, respectively, such that $Pr(\Q: (A \wedge \neg
\Int_1)) = 0$ and $Pr(\Q: (\Int_2 \wedge B)) = 0$.
In other words, $A \implies \Int_1$ and $\Int_2 \implies \neg B$, as
illustrated by the Karnaugh-Veitch diagrams in
Figure~\ref{Fig:ExampleInterpolSres}.

\section{Applications of generalized Craig interpolation to analysis of
probabilistic systems}
\label{Sec:Analysis}

\begin{figure}[t]
\centering
\resizebox{0.5\textwidth}{!}
{\input{example-MDP-pure.pstex_t}}
\caption{A simple MDP~$\MDP$.}
\label{Fig:ExampleMDP}
\end{figure}

In this section, we investigate the application of generalized Craig
interpolation to the symbolic analysis of probabilistic systems. We direct our
attention to two analysis goals, namely to \emph{probabilistic state
reachability} in Section~\ref{Subsec:Reachability} as well as to
\emph{probabilistic region stability} in Section~\ref{Subsec:Stability}.
As a system model, we consider finite-state Markov decision processes
(MDPs)~\cite{Bellman:MDP}. An MDP $\MDP = (\imath, S, {\it Act}, {\it ps}(\cdot,
\cdot, \cdot))$ is a finite-state system in which state changes are subject to
\emph{non-deterministic} selection among available actions followed by a
\emph{probabilistic} choice among potential successor states, while the
probability distribution of the latter choice depends on the selected action.
More precisely, $S$ is a finite set of states, $\imath \in S$ is the initial
state, ${\it Act}$ is a finite set of actions, and ${\it ps}(s,a,s')$ gives the
probability that $\MDP$ performs a transition step from $s \in S$ to $s' \in S$
under action $a \in {\it Act}$.
For an example, consider the simple MDP $\MDP$ from Figure~\ref{Fig:ExampleMDP}
where $\imath = i$, $S=\{i,f,e,s\}$, and ${\it Act} = \{a,b\}$. A transition
${\it ps}(z, {\it act}, z') = p > 0$ is indicated by an arrow from $z$ to $z'$
accompanied by action ${\it act}$ and by the corresponding transition
probability $p$. If two states are not connected by an arrow then the
corresponding transition probability is $0$, and if no action is specified then
that transition is feasible for all actions.
%
A probability measure of an MDP is well-defined only if considering a
particular scheduler $\sigma$ resolving the non-determinism. That is,
$\sigma$ schedules the action for the current state. Different such
schedulers $\sigma$ have been investigated in the literature, confer,
for instance, \cite{BaierHKH05}: $\sigma$ may select the next action
either in a deterministic or randomized fashion. In both cases,
$\sigma$ may have access to and thus base its selection on either the
current state only or the full system history.  In our scenarios, we
do not manipulate schedulers explicitly, but define the probability
measures obtained by worst-case deterministic schedulers achieving maximum or
minimum, depending on how the worst case is understood, probability of reaching
target states directly as the limit of a recursive function over $\Nat$. For
each $k\in\Nat$, the recursive function determines the maximum or minimum
probability of reaching target states within $k$~steps, as achieved by a
worst-case history-dependent scheduler. As a worst-case history-dependent
scheduler will always maximize or minimize the probability of reaching the
target within the remaining number of steps, its performance coincides with the
probabilities computed by a backward induction resolving non-deterministic
choices by taking the maximum or minimum, respectively, of the probability
values obtained from the next-lower recursion depth.

All experiments mentioned in this section were performed on a 1.83~GHz
Intel Core~2 Duo machine with 1~GByte physical memory running Linux.

\subsection{Interpolation-based probabilistic state reachability}
\label{Subsec:Reachability}

Let be given an MDP~$\MDP$ and a set of target states $\Target
\subseteq S$ in $\MDP$. With regard to \emph{probabilistic state
  reachability}, the goal is to compute the probability of reaching
the target states $\Target$ from the initial state $\imath$ under some
explicitly or implicitly (e.g., by an optimality condition) given
scheduler $\sigma$. In most applications, the target states are
considered to be \emph{bad}, for instance, to be fatal system errors,
such that one is faced with computing the \emph{worst-case}
probability of reaching the bad states, i.e.\ \emph{maximizing} the
reachability probability under each possible scheduler.  This maximum
probability $\MaxReach(\MDP,\Target)$ can be defined directly as
the limit of the maximum step-bounded probability of reaching the
target states as similarly shown by~\cite[Lemma~1]{FranzleHSCC2011},
i.e.\
\[
\MaxReach(\MDP,\Target) =
\lim_{k \to \infty} \MaxReach_{\MDP,\Target}^{k}(\imath)
\]
where
\[
\MaxReach_{\MDP,\Target}^{k}(s) = \left\{
\begin{array}{ll}
1 & ; s \in \Target\\
0 & ; s \notin \Target, k = 0\\
\max\limits_{a \in {\it Act}} \sum\limits_{s' \in S}  {\it ps}(s,a,s') \cdot
\MaxReach_{\MDP,\Target}^{k-1}(s') & ; s \notin \Target, k > 0\\
\end{array}\right.
\]
gives the maximum probability of reaching the target states from state $s \in
S$ within $k$ steps ($k \in \Nat$) under each possible scheduler.
For some threshold value $\theta \in [0,1]$, the \emph{safety verification
problem} is to decide whether the worst-case probability of reaching the bad
states is at most $\theta$, i.e.\ to decide whether
\begin{equation}
\label{Eq:ProblemSafety}
\MaxReach(\MDP,\Target) \leq \theta
\end{equation}
holds.

In previous work~\cite{FraHerTei08:SSMT,FraTeiEgg:JLAP,TeiEggFra:NAHS}, we have
established a symbolic \emph{falsification} procedure for above
problem~\ref{Eq:ProblemSafety}. Though this approach is based on SSMT, i.e.\ an
arithmetic extension of SSAT, and works for the more general class of
discrete-time probabilistic hybrid systems, which roughly are MDPs with
arithmetic-logical transition guards and actions, the same procedure restricted
to SSAT is applicable for finite-state MDPs. The key idea here is to adapt
\emph{bounded model checking} (BMC)~\cite{BiereEA:BMC} to the probabilistic case
by encoding step-bounded reachability as an SSAT problem: like in classical BMC,
the initial states, the transition relation, and the target states of an MDP
$\MDP$ are symbolically encoded by propositional formulae in CNF, namely by
$\Init(\vect{s})$, $\Trans(\vect{s}, \vect{nt}, \vect{pt}, \vect{s}')$, and
$\Target(\vect{s})$, respectively, where the propositional variable vector
$\vect{s}$ represents the system state before and $\vect{s}'$ after a transition
step. To keep track of the \emph{non-deterministic} and \emph{probabilistic}
selections of transitions in $\Trans(\vect{s}, \vect{nt}, \vect{pt},
\vect{s}')$, we further introduce propositional variables $\vect{nt}$ and
$\vect{pt}$ to encode non-deterministic selection among available actions and to
describe probabilistic choice of the successor state, respectively. Assignments
to these variables determine which of possibly multiple available transitions
departing from $\vect s$ is taken. In contrast to traditional BMC, all variables
are quantified: all state variables $\vect{s}$ and $\vect{s}'$ are existentially
quantified in the prefixes $\Q_{\vect{s}}$ and $\Q_{\vect{s}'}$. The
transition-selection variables $\vect{nt}$ encoding non-deterministic choice are
\emph{existentially quantified} by $\Q_{\vect{nt}}$, while the probabilistic
selector variables $\vect{pt}$ are bound by \emph{randomized quantifiers} in
$\Q_{\vect{pt}}$.\footnote{Non-deterministic branching of $n$ alternatives can
be represented by a binary tree of depth $\lceil\log_2 n\rceil$ and
probabilistic branching by a sequence of at most $n-1$ binary branches, yielding
$\lceil\log_2 n\rceil$ existential and $n-1$ randomized quantifiers,
respectively.} For the sake of clarity, let be $\vect{t} := \vect{nt} \cup
\vect{pt}$ and $\Q_{\vect{t}} := \Q_{\vect{nt}}\Q_{\vect{pt}}$.

According to~\cite[Proposition~1]{FraHerTei08:SSMT}, the maximum probability of
reaching the target states in $\MDP$ from the initial state $\imath$ within $k$
transition steps, i.e.\ $\MaxReach_{\MDP,\Target}^{k}(\imath)$, is equal to
the satisfaction probability
\begin{equation}
\label{Eq:lbReach}
lb_k :=
Pr\Big(
\Q(k):\Big(
\overbrace{\Init(\vect{s}_0) \wedge \bigwedge\nolimits_{i=1}^{k}
\Trans(\vect{s}_{i-1}, \vect{t}_i,
\vect{s}_i)}^{\text{states reachable within $k$ steps}}
\wedge
\overbrace{\left(\bigvee\nolimits_{i=0}^{k}\Target(\vect{s}_i)\right)}^{\text{
hit target states}}\Big)
\Big)
\end{equation}%
with $\Q(k):= \Q_{\vect{s}_0} \Q_{\vect{t}_1}\Q_{\vect{s}_1} \ldots
\Q_{\vect{s}_{k-1}}\Q_{\vect{t}_k}\Q_{\vect{s}_k}$.

Observe that each value $lb_k = \MaxReach_{\MDP,\Target}^{k}(\imath)$
can be computed by an SSAT solver and constitutes a lower bound of the
maximum reachability probability $\MaxReach(\MDP,\Target)$ due to
monotonicity of the chain
$\left(\MaxReach_{\MDP,\Target}^{k}(\imath)\right)_{k\in\Nat}$.  This
symbolic approach, called \emph{probabilistic bounded model checking}
(PBMC), is able to \emph{falsify} safety properties of
shape~\ref{Eq:ProblemSafety} once a value $lb_k > \theta$ is computed
for some $k$.

However, the development of a corresponding counterpart based on SSAT
that is able to compute \emph{upper} bounds $ub_k$ of the maximum
reachability probability $\MaxReach(\MDP,\Target)$ was left as an
open challenge. Such an approach would permit to \emph{verify} safety
properties of shape~\ref{Eq:ProblemSafety} once a value $ub_k \leq
\theta$ is computed for some $k$.

In the remainder of this section, we propose such a symbolic
\emph{verification} procedure for above problem~\ref{Eq:ProblemSafety}
by means of generalized Craig interpolation. This verification method
proceeds in two phases.  Phase~1 computes a symbolic representation of
an \emph{overapproximation of the backward reachable state set}, where
a state is backward reachable if it is the origin of a transition
sequence leading into $\Target$. Phase~1 can be integrated into PBMC, as
used to falsify the probabilistic safety property. Whenever such
falsification fails for a given step depth $k$, we apply generalized
Craig interpolation to the (just failed) PBMC proof to compute a
\emph{symbolic overapproximation of the backward reachable state set}
at depth $k$ and then proceed to PBMC at some higher depth $k'>k$. As
an alternative to the integration into PBMC, interpolants describing
the backward reachable state sets can be successively extended by
``stepping'' them by prepending another transition, as explained
below.  In either case, phase~1 ends when the backward reachable state
set becomes stable, in which case we have computed a symbolic
overapproximation of the whole backward reachable state set. In
phase~2, we construct an SSAT formula with parameter $k$ that forces
the system to \emph{stay within the backward reachable state set} for
$k$ steps. The maximum satisfaction probability of that SSAT formula
then gives an upper bound on the maximum probability of reaching the
target states. The rationale is that system runs leaving the backward
reachable state set will never reach the target states.

\paragraph{Phase~1.}
Given an SSAT encoding of an MDP $\MDP$ as above, the state-set predicate
$\B^{k}(\vect{s})$ for $k \in \Nat$ over state variables $\vect{s}$ is
inductively defined as
\begin{iteMize}{$\bullet$}
\item $\B^0(\vect{s}) := \Target(\vect{s})$, and
\item $\B^{k+1}(\vect{s}) := \B^{k}(\vect{s}) \vee \Int^{k+1}(\vect{s})$
\end{iteMize}
where
$\Int^{k+1}(\vect{s}_{j-1})$ is a generalized Craig interpolant for
\[
\Big(
\overbrace{\Trans(\vect{s}_{j-1}, \vect{t}_j, \vect{s}_j) \wedge
\B^k(\vect{s}_j)}^{=A},~~~~
\overbrace{\Init(\vect{s}_0) \wedge \bigwedge\nolimits_{i=1}^{j-1}
\Trans(\vect{s}_{i-1}, \vect{t}_i, \vect{s}_i)}^{=B}
\Big)
\]
with $j \geq 1$ with respect to SSAT formula
\begin{equation}
\label{Eq:CIReach}
\Q(j):
\Big(
\overbrace{\Init(\vect{s}_0) \wedge \bigwedge\nolimits_{i=1}^{j-1}
\Trans(\vect{s}_{i-1}, \vect{t}_i, \vect{s}_i)}^{\text{$j-1$ steps
``forward''}~~(=B)}
\wedge
\overbrace{\Trans(\vect{s}_{j-1}, \vect{t}_j, \vect{s}_j)
\wedge \B^k(\vect{s}_j)}^{\text{one step ``backward''}~~(=A)}\Big)~~.
\end{equation}%
Observe that each generalized Craig interpolant $\Int^{k+1}(\vect{s})$ can be
computed by interpolating S-resolution if we rewrite $\B^{k}(\vect{s})$ into
CNF, the latter being always possible in linear time by adding auxiliary
$V_A$-variables.
During computation of each $\Int^{k+1}(\vect{s})$, we take $I=\true$ in every
application of rule~\ref{Rule2:solution} such that $\B^{k}(\vect{s})$
overapproximates all system states backward reachable from target states within
$k$ steps due to Corollary~\ref{Cor:ControlSIntComp}.
Whenever $\B^{k}(\vect{s})$ has stabilized, i.e.\
\[
\B^{k+1}(\vect{s}) \implies \B^{k}(\vect{s})~~,
\]
we can be sure that $\BReach(\vect{s}) := \B^{k}(\vect{s})$ overapproximates all
backward reachable states.
It is obvious that $\B^{k}(\vect{s})$ finally stabilizes in the finite-state
case.

Note that parameter $j \geq 1$ can be chosen arbitrarily, i.e.\ the system may
execute any number of transitions until state $\vect{s}_{j-1}$ is reached since
this does not destroy the ``backward-overapproximating'' property of
$\B^{k+1}(\vect{s})$. The rationale of having parameter $j$ is the additional
freedom in constructing generalized interpolants since $j$ may influence the
shape of $\Int^{k+1}(\vect{s})$, as we will see in the example below.

We remark that phase~1 is a clean generalization of McMillan's
approach~\cite{McMillan03,McMillan:TACAS05}, the latter having unsatisfiability
of $A \wedge B$ as precondition in each iteration~$k$.\footnote{Instead of
overapproximating the backward reachable state set, McMillan's
scheme~\cite{McMillan03,McMillan:TACAS05} actually targets at forward
reachable states, which however makes no fundamental difference in the
non-probabilistic setting.}

\paragraph{Phase~2.}
Having symbolically described all backward reachable states by the predicate
$\BReach(\vect{s})$, \emph{upper} bounds $ub_k$ of the maximum probability
$\MaxReach(\MDP,\Target)$ of reaching the target states $\Target$ can now be
computed by SSAT solving applied to
\begin{equation}
\label{Eq:ubReach}
ub_k :=
Pr\Big(\Q(k):\Big(
\overbrace{\Init(\vect{s}_0) \wedge \bigwedge\nolimits_{i=1}^{k}
\Trans(\vect{s}_{i-1},
\vect{t}_i, \vect{s}_i)}^{\text{states reachable within $k$ steps}}
\wedge
\overbrace{\bigwedge\nolimits_{i=0}^{k}{\BReach}(\vect{s}_i)}^{\text{
stay in back-reach set}}\Big)\Big)~~.
\end{equation}%
First observe that the formula above excludes all system runs that leave the
set of backward reachable states. This is sound since leaving
$\BReach(\vect{s})$ means to never reach the $\Target(\vect{s})$ states.
Second, the system behavior becomes more and more constrained for increasing
$k$, i.e.\ the $ub_k$'s are monotonically decreasing. With regard to solving
problem~\ref{Eq:ProblemSafety}, the safety property $\MaxReach(\MDP,\Target)
\leq \theta$ is \emph{verified} by the procedure above once an upper bound $ub_k
\leq \theta$ is computed for some $k$.

\paragraph{Example.}
To illustrate the symbolic approach to probabilistic safety verification based
on generalized Craig interpolation, consider the simple MDP~$\MDP$ from
Figure~\ref{Fig:ExampleMDP} with $s$ being the only target state.

With regard to the symbolic encoding of $\MDP$, we introduce four
Boolean variables $i, f, e, s$ to describe the state space. The
literal $i$ means that $\MDP$ is in state $i$ while literal $\neg i$
expresses that $\MDP$ is not in $i$. The same holds analogously for
the other states. Note that, in order to encode valid system states,
we have to ensure that exactly one of the variables $i, f, e, s$ is
$\true$ in each time instant. The encoding of this constraint will be
explained later on.  The non-deterministic choice between actions $a$
and $b$ is encoded by a Boolean variable ${\it act}$ while action $a$
is represented by the positive literal ${\it act}$ and action $b$ by
the negative literal $\neg {\it act}$.  For the three probabilistic
choices in $\MDP$, we introduce three Boolean variables ${\it pi}$ for
the choice from $i$, ${\it pea}$ for the choice from $e$ under action
$a$, and ${\it peb}$ for the choice from $e$ under action $b$.  Recall
that all state variables as well as variables encoding
non-deterministic selection are existentially quantified while
variables describing probabilistic choices are bound by randomized
quantifiers. We thus obtain the corresponding quantifier prefixes
\[
\begin{array}{lll}
\Q_{\vect{s}}  & = & \exists i\ \exists f\ \exists e\ \exists s~~,\\
\Q_{\vect{t}}  & = & \exists {\it act}\ \R^{0.9} {\it pi}\ \R^{0.6} {\it pea}\
\R^{0.5} {\it peb}~~,\\
\Q_{\vect{s}'} & = & \exists i'\ \exists f'\ \exists e'\ \exists s'~~.\\
\end{array}
\]
The formulae in CNF representing the initial state and the target states are
specified by
\[
\Init(\vect{s}) = (i) \wedge (\neg f) \wedge (\neg e) \wedge (\neg s)
\text{~~~~and~~~~}
\Target(\vect{s}) = (s)~~,
\]
respectively.
To obtain the transition relation predicate, we encode each single transition
step. For instance, a step from state $e$ to $f$ under action $a$
can be encoded by the implication $(e \wedge {\it act} \wedge \neg {\it pea})
\implies f'$, the latter being equivalent to the clause $(\neg e \vee \neg {\it
act} \vee {\it pea} \vee f')$. The conjunction of all these clauses then encodes
the full system behavior symbolically. Since we represent each system state by
an own Boolean variable, as mentioned above, we need to enforce that exactly one
of the primed state variables, constituting the system state after the
transition step, carries value $\true$. This is simply achieved by the formula
in CNF
${\it exactly\_one}(i', f', e', s') =
(i' \vee f' \vee e' \vee s')
\wedge (\neg i' \vee \neg f')
\wedge (\neg i' \vee \neg e')
\wedge (\neg i' \vee \neg s')
\wedge (\neg f' \vee \neg e')
\wedge (\neg f' \vee \neg s')
\wedge (\neg e' \vee \neg s')$.
The transition relation predicate in CNF then is
\[
\begin{array}{llll}
\Trans(\vect{s}, \vect{t}, \vect{s}') & = & &
(\neg i \vee {\it pi} \vee f')
~~\wedge~~
(\neg i \vee \neg {\it pi} \vee e')\\
& & \wedge &
(\neg e \vee \neg {\it act} \vee {\it pea} \vee f')
~~\wedge~~
(\neg e \vee \neg {\it act} \vee \neg {\it pea} \vee s')\\
& & \wedge &
(\neg e \vee {\it act} \vee {\it peb} \vee s')
~~\wedge~~
(\neg e \vee {\it act} \vee \neg {\it peb} \vee i')\\
& & \wedge &
(\neg f \vee f')
~~\wedge~~
(\neg s \vee s')
~~\wedge~~
{\it exactly\_one}(i', f', e', s')~~.
\end{array}
\]

\begin{figure}[t]
\centering
\resizebox{0.49\textwidth}{!}
{\input{example-MDP-sets.pstex_t}}
\epsfxsize0.49\textwidth\epsffile{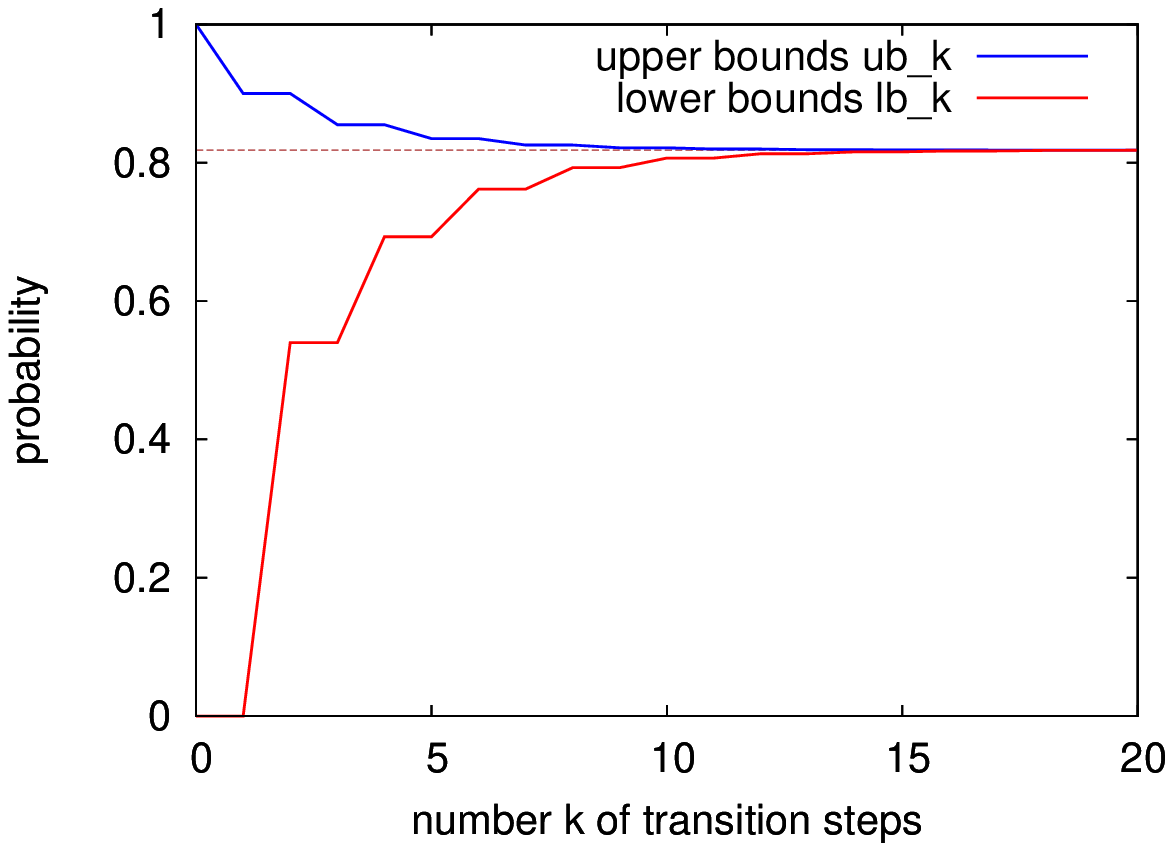}
\caption{Illustration of the computed state sets for MDP~$\MDP$ by the
generalized interpolation scheme~\ref{Eq:CIReach} with $j=3$ (left), and lower
bounds $lb_k$ and upper bounds $ub_k$ of the maximum probability of reaching
target state $s$ over number $k$ of transition steps computed by
schemes~\ref{Eq:lbReach} and \ref{Eq:ubReach}, respectively (right).}
\label{Fig:ResultsReachability}
\end{figure}

We are now interested in the maximum probability of reaching the target state
$s$ from the initial state $i$. Applying the PBMC scheme~\ref{Eq:lbReach}, we
are only able to compute lower bounds $lb_k$ of the maximum reachability
probability, for instance,
$lb_0 = lb_1 = 0$,
$lb_2 = lb_3 = 0.54$,
$lb_4 = lb_5 = 0.693$,
$\ldots$,
$lb_{20} = 0.817971$,
$\ldots$,
$lb_{100} = 0.81818181818181803208$.
The latter results were achieved by employing the SSMT solver
SiSAT\footnote{The SiSAT tool is available on
\url{http://sisat.gforge.avacs.org/}.}~\cite{TeiEggFra:NAHS} that provides a
convenient input language for specifying
probabilistic transition systems like MDPs. Unwinding of the system's transition
relation for increasing step bounds $k$, i.e.\ the construction of the SSAT
formulae specified by scheme~\ref{Eq:lbReach} in our context, is done fully
automatically. Furthermore, several algorithmic optimizations are exploited to
improve performance of the tool. Concerning runtime, all $100$ SSAT formulae
were solved within $37.05$~seconds, while computation of the first $20$ lower
bounds $lb_0$ to $lb_{20}$ just needed $370$~milliseconds. The highest
computation time for a single SSAT problem was obtained for $lb_{100}$, namely
$1.14$~seconds.
The evolution of the $lb_k$'s up to $k=20$ is presented graphically on the right
of Figure~\ref{Fig:ResultsReachability}. Given these results, one can suppose
that the lower bounds converge to and never exceed value $\nicefrac{9}{11} =
0.\overline{81}$. However, there is no mathematical guarantee for the latter
guess.

\begin{table}
\begin{tabular}{|c||c|c||c|c||c|c||c|}
\hline
$j$ & $\Int^1$ & $\B^1$ & $\Int^2$ & $\B^2$ & $\Int^3$ & $\B^3$ & $\BReach$\\
\hline\hline
$1$ & $\neg i$ & $\neg i \vee s$ & $\true$ & $\true$ & $\true$ & $\true$ &
$\true$\\
\hline
& $\{f,e,s\}$ & $\{f,e,s\}$ & $\{i,f,e,s\}$ & $\{i,f,e,s\}$ & $\{i,f,e,s\}$ &
$\{i,f,e,s\}$ & $\{i,f,e,s\}$\\
\hline\hline
$2$ & $\neg f$ & $\neg f \vee s$ & $\neg f$ & $\neg f \vee s$ & --- & --- &
$\neg f \vee s$\\
\hline
& $\{i,e,s\}$ & $\{i,e,s\}$ & $\{i,e,s\}$ & $\{i,e,s\}$ & --- & --- &
$\{i,e,s\}$\\
\hline\hline
$3$ & $\neg i \wedge \neg f$ & $\neg i \wedge \neg f \vee s$ & $\neg f$ & $\neg
f \vee s$ & $\neg f$ & $\neg f \vee s$ & $\neg f \vee s$\\
\hline
& $\{e,s\}$ & $\{e,s\}$ & $\{i,e,s\}$ & $\{i,e,s\}$ & $\{i,e,s\}$ & $\{i,e,s\}$
& $\{i,e,s\}$\\
\hline
\end{tabular}
\caption{Experimental results of applying the generalized interpolation
scheme~\ref{Eq:CIReach} on $\MDP$ from Figure~\ref{Fig:ExampleMDP} for different
values of parameter $j$. In addition to the formal presentation of the
predicates, the concrete state sets are given explicitly.}
\label{Table:ResultsBackreachSet}
\end{table}

To overcome this limitation, we first apply the generalized interpolation
scheme~\ref{Eq:CIReach} to compute an overapproximation of the backward
reachable state set. The latter then facilitates to compute upper bounds $ub_k$
of the maximum reachability probability by means of scheme~\ref{Eq:ubReach}.
In order to compute the generalized Craig interpolants
$\Int^{k+1}(\vect{s}_{j-1})$ automatically during solving the SSAT
formulae~\ref{Eq:CIReach}, we have implemented a simple DPLL-based SSAT solver
that integrates interpolating S-resolution.
As mentioned earlier, scheme~\ref{Eq:CIReach} allows freedom in choosing
parameter $j \geq 1$. This parameter permits to specify the number $j-1$ of
transition steps until system state $\vect{s}_{j-1}$ is reached, which is the
common state of formula parts $A$ and $B$.
The experimental results of applying the generalized interpolation
scheme~\ref{Eq:CIReach} on the MDP~$\MDP$ for different values of $j$ are shown
in Table~\ref{Table:ResultsBackreachSet}.

From the results of Table~\ref{Table:ResultsBackreachSet}, we observe
that the value of $j$ actually has an impact on the shape of the
resulting interpolants.  Let us consider the first interpolants
$\Int^1$ which overapproximate all states backward reachable in one
step. Clearly, the exact set of states backward reachable in one step
is $\{e,s\}$. For $j=1$, the overapproximated set $\{f,e,s\}$ computed
by the procedure is too coarse and actually contains a state which is
\emph{not} backward reachable at all, namely $f$. Though the set $\{i,e,s\}$ for
$j=2$ actually consists of backward reachable states only, it is not tight
enough as the initial state $i$ is backward reachable after two steps only.
For $j=3$, we achieved the precise set $\{e,s\}$.  Continuing the scheme for
$j=1$, $\Int^2$ and then $\Int^3$ become $\true$ meaning that the
overapproximated set of the backward reachable states $\BReach$ covers the whole
state space. Using this inconclusive result in scheme~\ref{Eq:ubReach} yields
only trivial upper bounds $ub_k = 1$ for all $k$.  With
regard to $j=2$, the interpolation process has stabilized after
computation of $\Int^2$. The resulting state set $\{i,e,s\}$ encoded
by $\BReach$ actually is the precise set of all backward reachable
states. Though $\Int^1$ was too coarse, this could be compensated in
the computation of $\Int^2$.  For $j=3$, we observe that all
generalized interpolants $\Int^1$, $\Int^2$, and $\Int^3$ describe the
corresponding backward reachable states accurately, thus leading to
the precise set of all backward reachable states. The computed state
sets for $j=3$ are illustrated on the left of
Figure~\ref{Fig:ResultsReachability}.  After having examined the
results above, it seems that the greater the value of $j$, i.e.\ the
more transition steps are performed, the more accurate the resulting
overapproximation of the backward reachable state set.

Concerning runtime, each generalized Craig interpolant was computed by the
interpolating DPLL-based SSAT solver within fractions of a second, where the
highest runtime of $36$~milliseconds was observed when computing $\Int^3$ for
$j=3$.

Having computed a symbolic representation $\BReach(\vect{s})$ of an
overapproximation of all backward reachable states, we are now able to compute
upper bounds $ub_k$ of the maximum reachability probability by means of
scheme~\ref{Eq:ubReach}, where we use $\BReach(\vect{s}) = \neg f \vee s$ as
obtained for $j=3$ as well as for $j=2$.
Again employing the SSMT tool SiSAT, some of the results  are
$ub_0 = 1$,
$ub_1 = ub_2 = 0.9$,
$ub_3 = ub_4 = 0.855$,
$ub_5 = ub_6 = 0.83475$,
$\ldots$,
$ub_{20}  = 0.818243$,
$\ldots$,
$ub_{100} = 0.81818181818181821948$.
Concerning runtime, all $100$ SSAT formulae were solved within $54.76$~seconds,
while computation of the first $20$ upper bounds $ub_0$ to $ub_{20}$ just needed
$400$~milliseconds. The highest computation time for a single SSAT problem was
obtained for $ub_{100}$, namely $1.77$~seconds.
The evolution of the $ub_k$'s up to $k=20$ is presented graphically on the right
of Figure~\ref{Fig:ResultsReachability}.

In addition to estimating the maximum reachability probability from below using
the PBMC scheme~\ref{Eq:lbReach}, we are now able to estimate the probability
also from above. In our example, we can \emph{safely} conclude that
\[
0.81818181818181803208 = lb_{100} \leq \MaxReach(\MDP,\{s\}) \leq ub_{100}
= 0.81818181818181821948
\]
holds where the difference $ub_{100} - lb_{100}$ is below $10^{-15}$. The
total computational effort for obtaining this precise result is about $92$ seconds. If
reduced accuracy suffices then runtime obviously improves. For instance, the fact
\[
0.817971 = lb_{20} \leq \MaxReach(\MDP,\{s\}) \leq ub_{20} = 0.818243
\]
with $ub_{20} - lb_{20} < 10^{-3}$ was deduced within one second.
With regard to the safety verification problem~\ref{Eq:ProblemSafety}, system
safety for each threshold value $\theta$ with $\theta < 0.817971$ or $\theta
\geq 0.818243$ is \emph{falsified} or \emph{verified}, respectively, within a
second.

With respect to competitive and more established methods based on value or
policy iteration, we observed that the runtime of our prototypic tool chain
does not compare favorably on the simple probabilistic reachability problem
above. For instance, the version~4.0.1 of the PRISM model
checker\footnote{More information can be found on
\url{http://www.prismmodelchecker.org/}.}~\cite{prism4:cav2011} solved the
problem in about $600$~milliseconds with a precision of $10^{-15}$ (returning
the result $0.8181818181818175$).

In spite of the above fact, we have identified two promising directions for
future research where probabilistic reachability analysis based on generalized
Craig interpolation may pay off:
\begin{enumerate}[(1)]
\item Embedding the same interpolation process into
  SSMT~\cite{FraHerTei08:SSMT}, i.e.\ an arithmetic extension of SSAT,
  renders the generalized Craig interpolation scheme~\ref{Eq:CIReach}
  \emph{directly} applicable to probabilistic \emph{hybrid}
  discrete-continuous systems, yielding a symbolic overapproximation
  of the backward reachable state set.  As for the finite state case,
  scheme~\ref{Eq:ubReach} then facilitates computing upper bounds of
  the reachability probability for hybrid systems by means of SSMT
  solving, just as already pursued when computing lower bounds according to
  the PBMC scheme~\ref{Eq:lbReach}~\cite{FraHerTei08:SSMT,
    TeiFra08:nonlinearSSMT,FraTeiEgg:JLAP,TeiEggFra:NAHS}.

It is important to remark that classical value or policy iteration procedures
are \emph{not} directly applicable in the hybrid state case but even after
finite-state abstraction, confer, for instance,
\cite{PHA:CAV10,FranzleHSCC2011}.

\item Due to its \emph{symbolic} nature, the analysis procedures based on
SSAT and SSMT support \emph{compact} representations of \emph{concurrent}
probabilistic (finite-state and hybrid) systems without an explicit construction
of the product automaton~\cite{TeiEggFra:NAHS}, the latter being of size
exponential in the number of parallel components.
This fact constitutes a strong argument that these symbolic procedures are able
to alleviate the state explosion problem, which arises necessarily when
applying explicit-state algorithms or methods based on finite-state abstraction
refinement.
\end{enumerate}

\subsection{Interpolation-based probabilistic region stability}
\label{Subsec:Stability}
In addition to probabilistic state reachability being investigated in the
previous section, we now address the problem of \emph{probabilistic
region stability}. For that purpose, we take into account the notion of
\emph{region stability} as introduced for non-probabilistic hybrid systems
by Podelski and Wagner
in~\cite{PodelskiW:FORMATS07,PodelskiW:HSCC07}. According to their definition, given
some set $R$ of states called \emph{region}, a (non-probabilistic) system is
called \emph{stable with respect to region $R$} iff for every infinite run
$\langle s_0, s_1, \ldots, s_i, \ldots \rangle$ of the system, i.e.\ for every
infinite sequence of states that follows the transition relation, there is some
point of time $i \geq 0$ such that from $i$ on the system remains in $R$
forever, i.e.\ $\exists i \geq 0\ \forall j \geq i: s_j \in R$.

Concerning the probabilistic case, several adaptations of region stability seem
feasible, some of which pose measurability problems. Our main concern
in this article being to identify potential application areas for
generalized Craig interpolation rather than to discuss semantic issues
of probabilistic stabilization, we do study a
simple notion of probabilistic region stability in the sequel which
circumvents measure-theoretic issues.
As for probabilistic state reachability, we aim at defining a reasonable
probability measure as the limit of the value of a recursive
function defining the corresponding step-bounded measures.
Intuitively, we consider finite run prefixes $\langle s_0, s_1, \ldots, s_i
\rangle$ such that from time point $i$ on the probabilistic system remains in
the given region forever under each possible future behavior, i.e.\ independent
of the non-deterministic and probabilistic choices the system will take.
The latter fact is guaranteed whenever the system has reached an invariance kernel
of the given region that can never be left.
The probability measure is then defined by the minimum probability of reaching
the maximal invariance kernel.

Formally, let be given an MDP~$\MDP$ and a set of states $\Region
\subseteq S$ called the \emph{stabilization region} or the
\emph{region} for short. An \emph{invariance kernel} $\Kernel
\subseteq \Region$ with respect to $\MDP$ is a set of states from
$\Region$ such that there is no transition from a state in $\Kernel$
to a state outside $\Kernel$, i.e.\ there does not exist a tuple $(z,
{\it act}, z') \in \Kernel \times {\it Act} \times \left(S \setminus
  \Kernel\right): {\it ps}(z, {\it act}, z') > 0$.  An invariance
kernel $\Kernel$ is called \emph{maximal} if adding any new states to
$\Kernel$ does not lead to an invariance kernel, i.e.\ each $\Kernel
\cup Z$ with $Z \subseteq \Region \setminus \Kernel$ and $Z \neq \emptyset$ is
not an invariance kernel.  Note that the maximal invariance kernel is
unique. The latter fact can be simply shown using the observation that
the set of all invariance kernels $\Kernel \subseteq \Region$ with
respect to $\MDP$ is closed under union.  Let $\Kernel^\ast \subseteq
\Region$ be the (unique) maximal invariance kernel with respect to
$\MDP$.  Then, the minimum probability $\MinStable(\MDP,\Region)$ that $\MDP$ is
\emph{stable with respect to $\Region$} is defined as the limit of the minimum
step-bounded probability of reaching the maximal invariance kernel
$\Kernel^\ast$, i.e.\
\[
\MinStable(\MDP,\Region) =
\lim_{k \to \infty} \MinReach_{\MDP,\Kernel^\ast}^{k}(\imath)
\]
where
\[
\MinReach_{\MDP,\Kernel^\ast}^{k}(s) = \left\{
\begin{array}{ll}
1 & ; s \in \Kernel^\ast\\
0 & ; s \notin \Kernel^\ast, k = 0\\
\min\limits_{a \in {\it Act}} \sum\limits_{s' \in S}  {\it ps}(s,a,s') \cdot
\MinReach_{\MDP,\Kernel^\ast}^{k-1}(s') & ; s \notin \Kernel^\ast, k > 0\\
\end{array}\right.
\]
gives the minimum probability of reaching $\Kernel^\ast$ from state $s \in
S$ within $k$ steps ($k \in \Nat$) under each possible scheduler.

When considering stabilization within $\Region$ as the \emph{desired property}
then the value of $\MinStable(\MDP,\Region)$ establishes the probability of
stabilizing in \emph{worst case}, i.e.\ under an optimal adversarial scheduler.
For some threshold value $\theta \in [0,1]$, the \emph{stability verification
problem} then is to decide whether this worst-case probability is at least
$\theta$, i.e.\ to decide whether
\begin{equation}
\label{Eq:ProblemStability}
\MinStable(\MDP,\Region) \geq \theta
\end{equation}
holds.

In what follows, we propose a symbolic \emph{verification}
procedure for above problem~\ref{Eq:ProblemStability}. In a first
phase, we compute a symbolic representation of an invariance
kernel by means of generalized Craig interpolation. The main idea here
is to iteratively eliminate states $z$ not belonging to an invariance
kernel from $\Region$ until a fixed point is reached.  Due to the use
of interpolation, the set of such states $z$ is overapproximated in
each iteration, meaning that potentially too many states are
removed. This implies that the resulting invariance kernel is
\emph{not} necessarily maximal. However, each invariance kernel can
be used for computing valid lower bounds of $\MinStable(\MDP,\Region)$. The
latter computation then is performed in a second phase by means of
SSAT-based bounded reachability checking. Once a lower bound $lb \geq \theta$ is
computed, property~\ref{Eq:ProblemStability} is verified.

\paragraph{Phase~1.}
Let be given an SSAT encoding of an MDP $\MDP$ as explained in
Section~\ref{Subsec:Reachability} as well as some propositional formula
$\Region(\vect{s})$
encoding the stabilization region $\Region$.
Then, the state-set predicate $\Reg^{k}(\vect{s})$ for $k \in \Nat$
over state variables $\vect{s}$ is inductively defined as
\begin{iteMize}{$\bullet$}
\item $\Reg^0(\vect{s}) := \Region(\vect{s})$, and
\item $\Reg^{k+1}(\vect{s}) := \Reg^{k}(\vect{s}) \wedge \neg
\Int^{k+1}(\vect{s})$
\end{iteMize}
where $\Int^{k+1}(\vect{s}_{j-1})$ is a generalized Craig interpolant for
\[
\Big(
\overbrace{\Trans(\vect{s}_{j-1}, \vect{t}_j, \vect{s}_j) \wedge \neg
\Reg^{k}(\vect{s}_j)}^{=A},~~~~
\overbrace{\Init(\vect{s}_0) \wedge \bigwedge\nolimits_{i=1}^{j-1}
\Trans(\vect{s}_{i-1}, \vect{t}_i, \vect{s}_i)}^{=B}
\Big)
\]
with $j \geq 1$ with respect to SSAT formula
\begin{equation}
\label{Eq:CIStable}
\Q(j):
\Big(
\overbrace{\Init(\vect{s}_0) \wedge \bigwedge\nolimits_{i=1}^{j-1}
\Trans(\vect{s}_{i-1}, \vect{t}_i, \vect{s}_i)}^{\text{$j-1$ steps
``forward''}~~(=B)}
\wedge
\overbrace{\Trans(\vect{s}_{j-1}, \vect{t}_j, \vect{s}_j)
\wedge \neg \Reg^{k}(\vect{s}_j)}^{\text{one step ``backward'' from $\neg
\Reg^{k}$}~~(=A)}\Big)~~.
\end{equation}%
Observe that each $\Int^{k+1}(\vect{s})$ can be computed by interpolating
S-resolution if we rewrite $\neg \Reg^{k}(\vect{s})$ into CNF, the latter being
always possible in linear time by adding auxiliary $V_A$-variables.
During computation of each $\Int^{k+1}(\vect{s})$, we take $I=\true$ in every
application of rule~\ref{Rule2:solution} such that $\Int^{k+1}(\vect{s})$
overapproximates all system states directly leading to the state set $\neg
\Reg^{k}(\vect{s})$ due to Corollary~\ref{Cor:ControlSIntComp}. As a
consequence, from each state in $\Reg^{k+1}(\vect{s}) = \Reg^{k}(\vect{s})
\wedge \neg \Int^{k+1}(\vect{s})$ it is infeasible to leave the set
$\Reg^{k}(\vect{s})$ in one step.
Whenever the chain $\Reg^{k}(\vect{s})$ has stabilized, i.e.\
\[
\Reg^{k}(\vect{s}) \implies \Reg^{k+1}(\vect{s})~~,
\]
it follows that $\Kernel(\vect{s}) := \Reg^{k}(\vect{s})$ is an invariance
kernel of $\Region(\vect{s})$ with respect to $\MDP$, i.e.\ once entered, the
system cannot leave the set $\Kernel(\vect{s})$.
Obviously, the chain $\Reg^{k}(\vect{s})$ eventually stabilizes in the finite-state case.

Similar to scheme~\ref{Eq:CIReach}, parameter $j \geq 1$ can be chosen
arbitrarily, i.e.\ the system may execute any number of transitions until state
$\vect{s}_{j-1}$ is reached since this does not destroy the
overapproximation property of $\Int^{k+1}(\vect{s})$. The presence of
parameter $j$ gives us additional freedom in constructing generalized
interpolants as $j$ may influence the shape of $\Int^{k+1}(\vect{s})$, as we
will see in the example below.

\paragraph{Phase~2.}
Having computed a symbolic representation $\Kernel(\vect{s})$ of a (not necessarily maximal)
invariance kernel $\Kernel$ with respect to $\MDP$, we now compute lower
bounds of the minimum probability $\MinStable(\MDP,\Region)$ of stabilizing
within $\Region$ by means of SSAT solving.
To this end, first observe that $\MinReach_{\MDP,\Kernel^\ast}^{k}(\imath)$
is monotonic in $k$ which implies that
$\MinReach_{\MDP,\Kernel^\ast}^{k}(\imath) \leq \MinStable(\MDP,\Region)$ for
each $k \in \Nat$.
Let $\Kernel^\ast$ be the unique maximal invariance kernel with respect to
$\MDP$. Then, $\Kernel \subseteq \Kernel^\ast$ since $\Kernel$ is an invariance
kernel and the maximal invariance kernel $\Kernel^\ast$ is unique. As a
consequence,
\[
\MinReach_{\MDP,\Kernel}^{k}(\imath) \leq
\MinReach_{\MDP,\Kernel^\ast}^{k}(\imath)
\]
for each $k \in \Nat$.
Summing up, each value of $\MinReach_{\MDP,\Kernel}^{k}(\imath)$ establishes
a lower bound of $\MinStable(\MDP,\Region)$.
In principle, $\MinReach_{\MDP,\Kernel}^{k}(\imath)$ can be reduced to an
SSAT formula similar to PBMC scheme~\ref{Eq:lbReach}. The difference, however,
is that we need to \emph{minimize} the satisfaction probability. The latter can
be achieved by a very similar SSAT encoding scheme that exploits
\emph{universal} quantifiers to resolve non-deterministic transition choices.
Universal quantifiers then aim at minimizing the satisfaction probability.
Though the SSMT solver SiSAT actually supports universal quantification,
confer~\cite{TeigeFraenzle09:ADHS,TeiEggFra:NAHS}, we instead stay within the
scope of the logic exposed in this article and rephrase minimum probabilistic
state reachability as a \emph{maximum probabilistic state avoidance problem} as
follows:
\[
\MaxAvoid_{\MDP,\Kernel}^{k}(s) = \left\{
\begin{array}{ll}
0 & ; s \in \Kernel\\
1 & ; s \notin \Kernel, k = 0\\
\max\limits_{a \in {\it Act}} \sum\limits_{s' \in S}  {\it ps}(s,a,s') \cdot
\MaxAvoid_{\MDP,\Kernel}^{k-1}(s') & ; s \notin \Kernel, k > 0\\
\end{array}\right.
\]
It then holds that
\[
\MinReach_{\MDP,\Kernel}^{k}(\imath) =
1 - \MaxAvoid_{\MDP,\Kernel}^{k}(\imath)
\]
which can be proven by straightforward induction over step bound $k$. In the
base cases, i.e.\ if $k = 0$ and $s \in \Kernel$ or $s \notin \Kernel$, the
statement is clear. Within the induction step, we exploit the property that
\[
\min\nolimits_i \sum\nolimits_j p_{i,j} \cdot P_{i,j} =
1 - \max\nolimits_i \sum\nolimits_j p_{i,j} \cdot (1-P_{i,j})
\]
is true for $0 \leq P_{i,j} \leq 1$ and $\sum_j p_{i,j} = 1$.

The problem of computing the value of $\MaxAvoid_{\MDP,\Kernel}^{k}(\imath)$
can be reduced to computing the maximum probability of satisfaction of the SSAT
formula
\[
\Phi_{\MDP,\Kernel}^{k}~~ =~~
\Q(k):\Big(
\overbrace{\Init(\vect{s}_0) \wedge \bigwedge\nolimits_{i=1}^{k}
\Trans(\vect{s}_{i-1},
\vect{t}_i, \vect{s}_i)}^{\text{states reachable within $k$ steps}}
\wedge
\overbrace{\bigwedge\nolimits_{i=0}^{k}{\neg \Kernel}(\vect{s}_i)}^{\text{
avoid invariance kernel}}\Big)~~.
\]
According to the definition of $\MaxAvoid_{\MDP,\Kernel}^{k}(\imath)$, the
propositional formula of $\Phi_{\MDP,\Kernel}^{k}$ describes all system runs
avoiding the invariance kernel $\Kernel$ for at least $k$ transition steps. That
is, all assignments encoding such latter runs yield satisfaction
probability~$1$, while assignments encoding runs that visit $\Kernel$ within the
first $k$ steps do not satisfy the propositional formula, thus leading to
satisfaction probability~$0$. As a consequence,
$\MaxAvoid_{\MDP,\Kernel}^{k}(\imath) = Pr\left(\Phi_{\MDP,\Kernel}^{k}\right)$.
Using above facts, we deduce the following relation
\[
\begin{array}{lcl}
1 - Pr\left(\Phi_{\MDP,\Kernel}^{k}\right) & = &
         1 - \MaxAvoid_{\MDP,\Kernel}^{k}(\imath)\\[1ex]
& = &    \MinReach_{\MDP,\Kernel}^{k}(\imath)\\[1ex]
& \leq & \MinReach_{\MDP,\Kernel^\ast}^{k}(\imath)\\[1ex]
& \leq & \MinStable(\MDP,\Region)~~.
\end{array}
\]
This finally enables us to compute lower bounds $lb_k$ of
$\MinStable(\MDP,\Region)$ using the scheme
\begin{equation}
\label{Eq:lbReachKernel}
lb_k := 1 - Pr\left(\Phi_{\MDP,\Kernel}^{k}\right)~~,
\end{equation}%
the latter being addressed by SSAT solving.
Note that the system behavior encoded by $\Phi_{\MDP,\Kernel}^{k}$ becomes more
and more constrained for increasing $k$ such that the satisfaction probabilities
$Pr\left(\Phi_{\MDP,\Kernel}^{k}\right)$ are monotonically decreasing. This in
turn means that the $lb_k$'s are monotonically increasing. With regard to
solving the stability verification problem~\ref{Eq:ProblemStability}, the
desired property $\MinStable(\MDP,\Region) \geq \theta$ is \emph{verified}
by the procedure above once a lower bound $lb_k \geq \theta$ is computed for
some $k$.

\paragraph{Example.}
To illustrate the symbolic approach to probabilistic region stability based
on generalized Craig interpolation, again consider the simple MDP~$\MDP$ from
Figure~\ref{Fig:ExampleMDP} where the symbolic representation of the region is
given by $\Region(\vect{s}) = \neg f$. That is, the region in which $\MDP$
should stabilize consists of the states $i$, $e$, and $s$.
The symbolic SSAT encoding of $\MDP$ being introduced in the example of
Section~\ref{Subsec:Reachability} is reused in the following.

\begin{table}
\begin{tabular}{|c||c|c||c|c||c|c||c|}
\hline
$j$ & $\Int^1$ & $\Reg^1$ & $\Int^2$ & $\Reg^2$ & $\Kernel$\\
\hline\hline
$1$ & $\true$ & $\false$ & $\true$ & $\false$ & $\false$\\
\hline
& $\{i,f,e,s\}$ & $\emptyset$ & $\{i,f,e,s\}$ & $\emptyset$ & $\emptyset$\\
\hline\hline
$2$ & $\true$ & $\false$ & $\true$ & $\false$ & $\false$\\
\hline
& $\{i,f,e,s\}$ & $\emptyset$ & $\{i,f,e,s\}$ & $\emptyset$ & $\emptyset$\\
\hline\hline
$3$ & $\neg s$ & $\neg f \wedge s$ & $\neg s$ & $\neg f \wedge s$ & $\neg f
\wedge s$\\
\hline
& $\{i,f,e\}$ & $\{s\}$ & $\{i,f,e\}$ & $\{s\}$ & $\{s\}$\\
\hline\hline
$4$ & $\neg s$ & $\neg f \wedge s$ & $\neg s$ & $\neg f \wedge s$ & $\neg f
\wedge s$\\
\hline
& $\{i,f,e\}$ & $\{s\}$ & $\{i,f,e\}$ & $\{s\}$ & $\{s\}$\\
\hline
\end{tabular}
\caption{Experimental results of applying the generalized interpolation
scheme~\ref{Eq:CIStable} on $\MDP$ from Figure~\ref{Fig:ExampleMDP} for
different values of parameter $j$. In addition to the symbolic
representations computed by interpolation, the concrete state sets
represented by these predicates are stated explicitly.}
\label{Table:ResultsInvarianceKernel}
\end{table}

\begin{figure}[t]
\centering
\resizebox{0.49\textwidth}{!}
{\input{example-MDP-sets-stability.pstex_t}}
\epsfxsize0.49\textwidth\epsffile{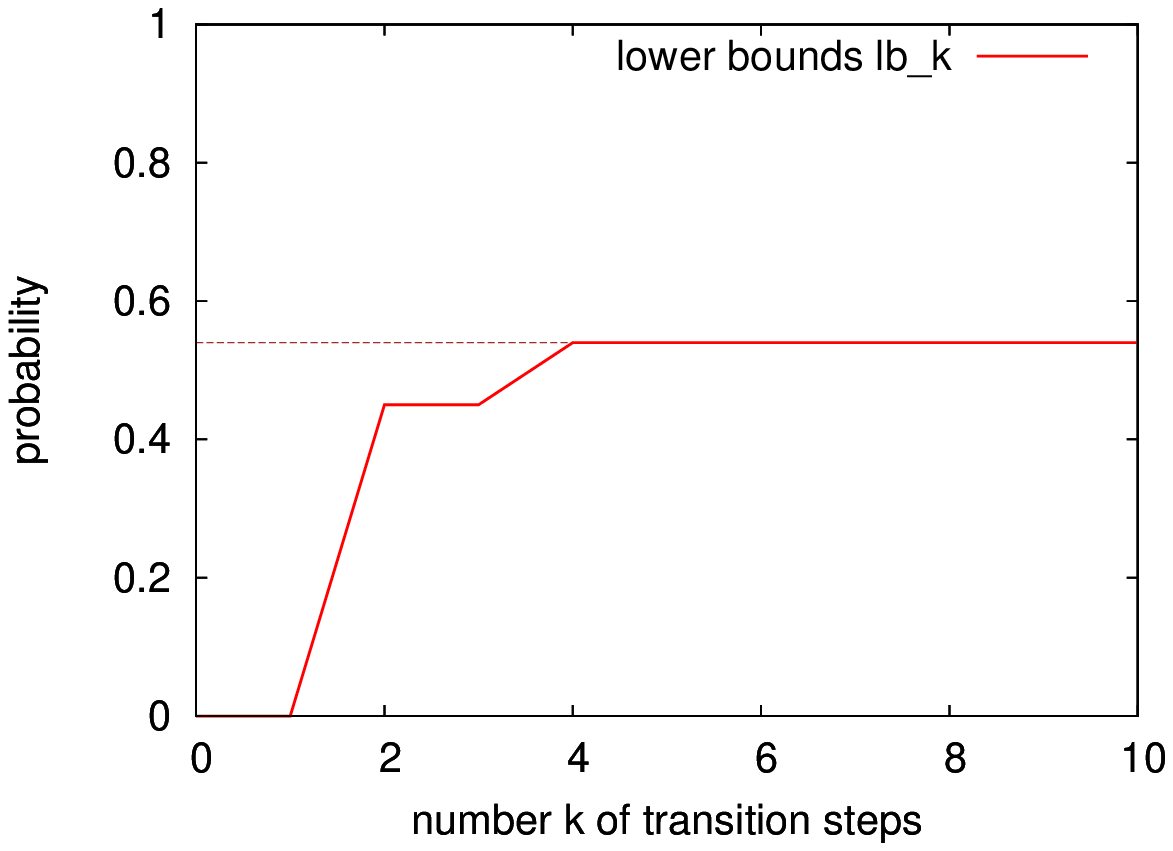}
\caption{Illustration of the computed state sets for MDP~$\MDP$ by the
generalized interpolation scheme~\ref{Eq:CIStable} with $j \in \{3,4\}$ (left),
and lower bounds $lb_k$ of the minimum probability of reaching the invariance
kernel $\Kernel=\{s\}$ over number $k$ of transition steps computed by
scheme~\ref{Eq:lbReachKernel} (right).}
\label{Fig:ResultsStability}
\end{figure}

We are first interested in computing an invariance kernel $\Kernel \subseteq
\Region(\vect{s})$ with respect to $\MDP$ by means of the generalized Craig
interpolation scheme~\ref{Eq:CIStable}. To cope with the latter scheme
automatically, we employ the simple interpolating DPLL-based SSAT solver
mentioned in Section~\ref{Subsec:Reachability}.
The results of these experiments for different values of $j$ are shown
in Table~\ref{Table:ResultsInvarianceKernel}.
It is not hard to see that the unique maximal invariance kernel $\Kernel^\ast$
consists of the state $s$ only.
Recall that each interpolant $\Int^{k+1}$ overapproximates all system states
directly leading to the state set $\neg \Reg^{k}$. When setting parameter $j$ to
value $1$ or $2$, we observe that interpolant $\Int^1=\true$ is too coarse since
it includes the whole state space. This causes the trivial invariance kernel
$\Kernel=\false$ representing the empty set. For choices $j=3$ and $j=4$,
however, $\Int^1=\neg s$ describes the exact set of states which lead to $\neg
\Reg^0 = \neg \Region$. Finally, the non-trivial invariance kernel $\Kernel =
\neg f \wedge s$ consisting of state $s$ only is computed. Note that $\Kernel$
actually is the maximal invariance kernel.
The computed state sets for $j \in \{3,4\}$ are illustrated on the left of
Figure~\ref{Fig:ResultsStability}.

These results confirm the observation made from the experiments of
Section~\ref{Subsec:Reachability}, namely that the greater the value of
$j$, i.e.\ the more transition steps are performed, the more accurate the
resulting overapproximations.
Concerning runtime, each generalized Craig interpolant was computed by the
interpolating DPLL-based SSAT solver within fractions of a second, where the
highest runtime of $88$~milliseconds was observed when computing $\Int^2$ for
$j=4$.

Having computed an invariance kernel $\Kernel(\vect{s}) \subseteq
\Region(\vect{s})$ with respect to $\MDP$, we are now able to compute
lower bounds $lb_k$ of the minimum probability that $\MDP$ is stable with
respect to $\Region$ by means of scheme~\ref{Eq:lbReachKernel}, where we use
$\Kernel(\vect{s}) = \neg f \wedge s$ as obtained for $j \in \{3,4\}$.
Employing the SSMT tool SiSAT, some of the results are
$lb_0 = lb_1 = 0$,
$lb_2 = lb_3 = 0.45$,
$lb_4 = lb_5 = 0.54$,
$\ldots$,
$lb_{100} = 0.54$.
Concerning runtime, all $100$ SSAT formulae were solved within $88.16$~seconds,
while computation of the first $20$ lower bounds $lb_0$ to $lb_{20}$ just needed
$600$~milliseconds. The highest computation time for a single SSAT problem was
obtained for $lb_{100}$, namely $2.91$~seconds.
The evolution of the $lb_k$'s up to $k=10$ is presented graphically on the right
of Figure~\ref{Fig:ResultsStability}.
With regard to the stability verification problem~\ref{Eq:ProblemStability},
the desired property $\MinStable(\MDP,\Region) \geq \theta$ is
\emph{verified} for each threshold value $\theta \leq 0.54$ within a second.

Concerning competitive approaches, we remark that the probabilistic model
checking tool PRISM~4.0.1~\cite{prism4:cav2011} is also able to deal with
probabilistic region stability of MDPs by means of path
operators.\footnote{Confer
\url{http://www.prismmodelchecker.org/manual/PropertySpecification/ThePOperator}
for more detailed information.}
To determine the value of $\MinStable(\MDP,\Region)$ for the example above,
we used the specification {\tt Pmin=?\,[F\,P>=1\,[G\,(!f)]]} meaning that we are
interested in the minimum probability ({\tt Pmin=?}) that finally ({\tt F}) the
system satisfies almost surely ({\tt P>=1}) the property that globally ({\tt G})
state $f$ is never visited ({\tt !f}). PRISM solved the problem in
$644$~milliseconds returning the result $0.54$.

As discussed for the case of probabilistic state reachability at the end of
Section~\ref{Subsec:Reachability}, we are also confident that the presented
approach to probabilistic region stability based on generalized
Craig interpolation becomes beneficial when adapted to probabilistic
\emph{hybrid} systems, where the classical procedures are not
directly applicable. Furthermore, a particular pay-off is expected when dealing with
\emph{concurrent} probabilistic systems owing to the \emph{symbolic} nature of
the interpolation-based technique.

\section{Conclusion and future work}
\label{Sec:Conclusion}

In this article, we elaborated on the idea of Craig interpolation for
stochastic Boolean satisfiability. In consideration of the
difficulties that arise in this stochastic extension of the
propositional satisfiability problem, we first proposed a suitable
definition of a generalized Craig interpolant and then presented an
algorithm for automatically computing such interpolants. For the
latter purpose, we enhanced the SSAT resolution calculus by
corresponding rules for the construction of generalized Craig
interpolants. We furthermore demonstrated two applications of
generalized Craig interpolation as a means of automated analysis of
probabilistic finite-state systems.

We first considered probabilistic state reachability. The resulting procedure is
able to verify probabilistic safety requirements of the form ``the worst-case
probability of reaching undesirable system states is at most some given safety
threshold''. This complements the existing SSAT-based probabilistic bounded
model checking approach, which mechanizes falsification of such
safety properties.
As a second application, we gave attention to probabilistic region stability
and presented a symbolic technique for verifying stability properties like ``the
worst-case probability that the system stabilizes within some given region is
at least some given safety threshold''.

For future work, we are particularly interested in the adaptation of generalized
Craig interpolation to SSMT, i.e.\ the extension of SSAT with arithmetic
theories. One of the most challenging issues here will be the enhancement of the
SSAT resolution calculus as well as the corresponding rules for the construction
of generalized interpolants in order to deal with SSMT problems.
The ability of computing generalized Craig interpolants for SSMT would lift
the interpolation schemes~\ref{Eq:CIReach} and \ref{Eq:CIStable} to SSMT
problems, thus establishing symbolic verification approaches to probabilistic
state reachability and to probabilistic region stability for discrete-time
probabilistic hybrid systems.
We are confident that such symbolic procedures will prove beneficial within
the analysis of probabilistic hybrid systems, in particular when systems with a
high degree of concurrency are considered.

\section*{Acknowledgement}
The authors wish to acknowledge fruitful discussions with the researchers in
the AVACS project as well as in the MoVeS project, in particular with Andreas
Eggers.
Furthermore, we would like to thank the anonymous reviewers for their advice on
how to enhance readability of the article.

\bibliography{ref_TT}

\newcommand{\etalchar}[1]{$^{#1}$}
\begin{thebibliography}{BHvMW09}

\bibitem[BCCZ99]{BiereEA:BMC}
Armin Biere, Alessandro Cimatti, Edmund~M. Clarke, and Yunshan Zhu.
\newblock Symbolic model checking without {BDD}s.
\newblock In Rance Cleaveland, editor, {\em Proceedings of the 5th
  International Conference on Tools and Algorithms for Construction and
  Analysis of Systems, TACAS '99}, volume 1579 of {\em Lecture Notes in
  Computer Science}, pages 193--207. Springer, 1999.

\bibitem[Bel57]{Bellman:MDP}
Richard Bellman.
\newblock A {M}arkovian decision process.
\newblock {\em Journal of Mathematics and Mechanics}, 6(5):679--684, 1957.

\bibitem[BHKH05]{BaierHKH05}
Christel Baier, Holger Hermanns, Joost-Pieter Katoen, and Boudewijn~R.
  Haverkort.
\newblock Efficient computation of time-bounded reachability probabilities in
  uniform continuous-time {M}arkov decision processes.
\newblock {\em Theor. Comput. Sci.}, 345(1):2--26, 2005.

\bibitem[BHvMW09]{HandbookOfSAT2009}
Armin Biere, Marijn J.~H. Heule, Hans van Maaren, and Toby Walsh, editors.
\newblock {\em Handbook of Satisfiability}, volume 185 of {\em Frontiers in
  Artificial Intelligence and Applications}.
\newblock IOS Press, February 2009.

\bibitem[BKF95]{BuningKF95}
Hans~Kleine B{\"u}ning, Marek Karpinski, and Andreas Fl{\"o}gel.
\newblock Resolution for quantified {B}oolean formulas.
\newblock {\em Inf. Comput.}, 117(1):12--18, 1995.

\bibitem[BS06]{BalafoutisS06}
Thanasis Balafoutis and Kostas Stergiou.
\newblock Algorithms for stochastic {CSP}s.
\newblock In Fr{\'e}d{\'e}ric Benhamou, editor, {\em Proceedings of the 12th
  International Conference on Principles and Practice of Constraint Programming
  (CP 2006)}, volume 4204 of {\em Lecture Notes in Computer Science}, pages
  44--58. Springer, 2006.

\bibitem[BS07]{BordeauxS07}
Lucas Bordeaux and Horst Samulowitz.
\newblock On the stochastic constraint satisfaction framework.
\newblock In {\em Proceedings of the 2007 ACM Symposium on Applied Computing
  (SAC)}, pages 316--320. ACM, 2007.

\bibitem[BSST09]{BSST09HBSAT}
Clark Barrett, Roberto Sebastiani, Sanjit~A. Seshia, and Cesare Tinelli.
\newblock Satisfiability modulo theories.
\newblock In Biere et~al. \cite{HandbookOfSAT2009}, chapter~26, pages 825--885.

\bibitem[Cra57]{Craig57}
William Craig.
\newblock Linear reasoning. a new form of the {H}erbrand-{G}entzen theorem.
\newblock {\em J. Symb. Log.}, 22(3):250--268, 1957.

\bibitem[DLL62]{DavisEA:DPLL62}
Martin Davis, George Logemann, and Donald~W. Loveland.
\newblock A machine program for theorem-proving.
\newblock {\em Commun. ACM}, 5(7):394--397, 1962.

\bibitem[DP60]{DavisPutnam}
Martin Davis and Hilary Putnam.
\newblock A computing procedure for quantification theory.
\newblock {\em Journal of the ACM}, 7(3):201--215, 1960.

\bibitem[FHH{\etalchar{+}}11]{FranzleHSCC2011}
Martin Fr\"{a}nzle, Ernst~Moritz Hahn, Holger Hermanns, Nicol\'{a}s Wolovick,
  and Lijun Zhang.
\newblock Measurability and safety verification for stochastic hybrid systems.
\newblock In {\em Proceedings of the 14th International Conference on Hybrid
  Systems: Computation and Control (HSCC 2011)}, pages 43--52, New York, NY,
  USA, 2011. ACM.

\bibitem[FHT08]{FraHerTei08:SSMT}
Martin Fr{\"a}nzle, Holger Hermanns, and Tino Teige.
\newblock Stochastic satisfiability modulo theory: A novel technique for the
  analysis of probabilistic hybrid systems.
\newblock In Magnus Egerstedt and Bud Mishra, editors, {\em Proceedings of the
  11th International Conference on Hybrid Systems: Computation and Control
  (HSCC 2008)}, volume 4981 of {\em Lecture Notes in Computer Science}, pages
  172--186. Springer, 2008.

\bibitem[FTE10]{FraTeiEgg:JLAP}
Martin Fr{\"a}nzle, Tino Teige, and Andreas Eggers.
\newblock Engineering constraint solvers for automatic analysis of
  probabilistic hybrid automata.
\newblock {\em Journal of Logic and Algebraic Programming}, 79(7):436--466,
  2010.

\bibitem[KNP11]{prism4:cav2011}
Marta Kwiatkowska, Gethin Norman, and David Parker.
\newblock Prism 4.0: Verification of probabilistic real-time systems.
\newblock In Ganesh Gopalakrishnan and Shaz Qadeer, editors, {\em Proceedings
  of the 23rd International Conference on Computer Aided Verification (CAV
  2011)}, volume 6806 of {\em Lecture Notes in Computer Science}, pages
  585--591. Springer, 2011.

\bibitem[Lit99]{littman99initial}
Michael~L. Littman.
\newblock Initial experiments in stochastic satisfiability.
\newblock In {\em Proceedings of the 16th National Conference on Artificial
  Intelligence}, pages 667--672, 1999.

\bibitem[LMP01]{littman01stochastic}
Michael~L. Littman, Stephen~M. Majercik, and Toniann Pitassi.
\newblock Stochastic {B}oolean satisfiability.
\newblock {\em Journal of Automated Reasoning}, 27(3):251--296, 2001.

\bibitem[Maj04]{Majercik04Nonchronological}
Stephen~M. Majercik.
\newblock Nonchronological backtracking in stochastic {B}oolean satisfiability.
\newblock In {\em 16th IEEE International Conference on Tools with Artificial
  Intelligence (ICTAI 2004)}, pages 498--507. IEEE Computer Society, 2004.

\bibitem[Maj09]{Maj09HBSAT}
Stephen~M. Majercik.
\newblock Stochastic {B}oolean satisfiability.
\newblock In Biere et~al. \cite{HandbookOfSAT2009}, chapter~27, pages 887--925.

\bibitem[McM03]{McMillan03}
Kenneth~L. McMillan.
\newblock Interpolation and {SAT}-based model checking.
\newblock In Warren A.~Hunt Jr. and Fabio Somenzi, editors, {\em Proceedings of
  the 15th International Conference on Computer Aided Verification (CAV 2003)},
  volume 2725 of {\em Lecture Notes in Computer Science}, pages 1--13.
  Springer, 2003.

\bibitem[McM05]{McMillan:TACAS05}
Kenneth~L. McMillan.
\newblock Applications of {Craig} interpolants in model checking.
\newblock In Nicolas Halbwachs and Lenore~D. Zuck, editors, {\em Proceedings of
  the 11th International Conference on Tools and Algorithms for the
  Construction and Analysis of Systems (TACAS 2005)}, volume 3440 of {\em
  Lecture Notes in Computer Science}, pages 1--12. Springer, 2005.

\bibitem[ML98]{majercik98maxplan}
Stephen~M. Majercik and Michael~L. Littman.
\newblock {MAXPLAN}: A new approach to probabilistic planning.
\newblock In {\em Proceedings of the Fourth International Conference on
  Artificial Intelligence Planning Systems}, pages 86--93. AAAI, 1998.

\bibitem[ML03]{majercik03contingent}
Stephen~M. Majercik and Michael~L. Littman.
\newblock Contingent planning under uncertainty via stochastic satisfiability.
\newblock {\em Artificial Intelligence Special Issue on Planning with
  Uncertainty and Incomplete Information}, 147(1-2):119--162, 2003.

\bibitem[Pap85]{Papadimitriou85}
Christos~H. Papadimitriou.
\newblock Games against nature.
\newblock {\em J. Comput. Syst. Sci.}, 31(2):288--301, 1985.

\bibitem[Pud97]{Pudlak97}
Pavel Pudl\'{a}k.
\newblock Lower bounds for resolution and cutting plane proofs and monotone
  computations.
\newblock {\em Journal of Symbolic Logic}, 62(3):981--998, September 1997.

\bibitem[PW07a]{PodelskiW:FORMATS07}
Andreas Podelski and Silke Wagner.
\newblock Region stability proofs for hybrid systems.
\newblock In Jean-Fran\c{c}ois Raskin and P.~S. Thiagarajan, editors, {\em
  Proceedings of the 5th International Conference on Formal Modeling and
  Analysis of Timed Systems (FORMATS 2007)}, volume 4763 of {\em Lecture Notes
  in Computer Science}, pages 320--335. Springer, 2007.

\bibitem[PW07b]{PodelskiW:HSCC07}
Andreas Podelski and Silke Wagner.
\newblock A sound and complete proof rule for region stability of hybrid
  systems.
\newblock In Alberto Bemporad, Antonio Bicchi, and Giorgio~C. Buttazzo,
  editors, {\em Proceedings of the 10th International Workshop on Hybrid
  Systems: Computation and Control (HSCC 2007)}, volume 4416 of {\em Lecture
  Notes in Computer Science}, pages 750--753. Springer, 2007.

\bibitem[Rob65]{Robinson:Resolution65}
John~Alan Robinson.
\newblock A machine-oriented logic based on the resolution principle.
\newblock {\em J. ACM}, 12(1):23--41, 1965.

\bibitem[TEF11]{TeiEggFra:NAHS}
Tino Teige, Andreas Eggers, and Martin Fr{\"a}nzle.
\newblock Constraint-based analysis of concurrent probabilistic hybrid systems:
  An application to networked automation systems.
\newblock {\em Nonlinear Analysis: Hybrid Systems}, 5(2):343--366, 2011.

\bibitem[TF08]{TeiFra08:nonlinearSSMT}
Tino Teige and Martin Fr{\"a}nzle.
\newblock Stochastic satisfiability modulo theories for non-linear arithmetic.
\newblock In Laurent Perron and Michael~A. Trick, editors, {\em Proceedings of
  the 5th International Conference on Integration of AI and OR Techniques in
  Constraint Programming for Combinatorial Optimization Problems (CPAIOR
  2008)}, volume 5015 of {\em Lecture Notes in Computer Science}, pages
  248--262. Springer, 2008.

\bibitem[TF09]{TeigeFraenzle09:ADHS}
Tino Teige and Martin Fr{\"a}nzle.
\newblock Constraint-based analysis of probabilistic hybrid systems.
\newblock In Alessandro Giua, Cristian Mahulea, Manuel Silva, and Janan
  Zaytoon, editors, {\em Proceedings of the 3rd IFAC Conference on Analysis and
  Design of Hybrid Systems}, pages 162--167. IFAC, 2009.

\bibitem[TF10]{TeiFrae:LPAR17}
Tino Teige and Martin Fr{\"a}nzle.
\newblock Resolution for stochastic {B}oolean satisfiability.
\newblock In Christian~G. Ferm{\"u}ller and Andrei Voronkov, editors, {\em
  Proceedings of the 17th International Conference on Logic for Programming,
  Artificial Intelligence, and Reasoning (LPAR-17)}, volume 6397 of {\em
  Lecture Notes in Computer Science}, pages 625--639. Springer, 2010.

\bibitem[Wal02]{walsh02stochastic}
Toby Walsh.
\newblock Stochastic constraint programming.
\newblock In Frank van Harmelen, editor, {\em Proceedings of the 15th European
  Conference on Artificial Intelligence (ECAI 2002)}, pages 111--115. IOS
  Press, 2002.

\bibitem[ZSR{\etalchar{+}}10]{PHA:CAV10}
Lijun Zhang, Zhikun She, Stefan Ratschan, Holger Hermanns, and Ernst~Moritz
  Hahn.
\newblock Safety verification for probabilistic hybrid systems.
\newblock In Tayssir Touili, Byron Cook, and Paul Jackson, editors, {\em
  Proceedings of the 22nd International Conference on Computer Aided
  Verification, CAV 2010}, volume 6174 of {\em Lecture Notes in Computer
  Science}, pages 196--211. Springer, 2010.

\end{thebibliography}

\bibliographystyle{alpha}




\end{document}